\documentclass[aps,twocolumn,showpacs,preprintnumbers,nofootinbib,prd,superscriptaddress,groupedaddress,10pt]{revtex4-1}
\usepackage[utf8]{inputenc}
\usepackage{graphicx,amssymb,amsmath,amsthm,amsfonts,epsfig,times,natbib}

\usepackage[usenames,dvipsnames]{color}
\usepackage{epstopdf}
\usepackage{amsmath,amssymb}
\usepackage{aas_macros}
\usepackage{tensor}
\usepackage{mathtools}
\usepackage{amsbsy}
\usepackage{bm,url}
\usepackage[linktocpage]{hyperref}

\usepackage[scaled]{beramono}
\usepackage[T1]{fontenc}

\definecolor{oxfordblue}{rgb}{0.0, 0.13, 0.28}
\definecolor{burgundy}{rgb}{0.5, 0.0, 0.13}
\definecolor{darkolivegreen}{rgb}{0.33, 0.42, 0.18}
\definecolor{darkblue}{rgb}{0,0,0.5}
\definecolor{richcarmine}{rgb}{0.84, 0.0, 0.25}
\definecolor{darkblue}{rgb}{0,0,0.5}
\definecolor{venetianred}{rgb}{0.78, 0.03, 0.08}
\definecolor{skobeloff}{rgb}{0.0, 0.48, 0.45}
\hypersetup{colorlinks=true, citecolor=darkblue, linkcolor=darkblue,
urlcolor = darkblue}

\newcommand{\dd}{\ensuremath{\mathrm{d}}}
\newcommand{\diff}[2]{\ensuremath{\frac{\dd {#1}}{\dd {#2}}}}
\newcommand{\msun}{\ensuremath{M_\odot}}

\def\nn{\nonumber}

\newcommand{\ben}{\begin{enumerate}}
\newcommand{\een}{\end{enumerate}}

\def\be{\begin{equation}}
\def\ee{\end{equation}}
\def\bea{\begin{eqnarray}}
\def\eea{\end{eqnarray}}
\def\nn{\nonumber}
\newcommand{\beq}{\begin{eqnarray}}
\newcommand{\eeq}{\end{eqnarray}} 
\newcommand{\ba}{\begin{align}}
\newcommand{\ea}{\end{align}}

\begin{document}

\title{
Gravitational wave searches for ultralight bosons with LIGO and LISA}

\author{
Richard Brito$^{1,\,*}$,
Shrobana Ghosh$^{2}$,
Enrico Barausse$^{3}$,
Emanuele Berti$^{2,4}$,
Vitor Cardoso$^{4,5}$,
Irina Dvorkin$^{3,6}$,
Antoine Klein$^{3}$,
Paolo Pani$^{7,4}$
}
\affiliation{${^1}$ Max Planck Institute for Gravitational Physics (Albert Einstein Institute), Am M\"{u}hlenberg 1, Potsdam-Golm, 14476, Germany}
\affiliation{${^2}$ Department of Physics and Astronomy, The University of 
Mississippi, University, MS 38677, USA}
\affiliation{${^3}$ Institut d'Astrophysique de Paris, Sorbonne
  Universit\'es, UPMC Univ Paris 6 
  \& CNRS, UMR 7095, 98 bis bd Arago, 75014 Paris, France}
\affiliation{${^4}$ CENTRA, Departamento de F\'isica, Instituto Superior
T\'ecnico, Universidade de Lisboa, Avenida Rovisco Pais 1,
1049 Lisboa, Portugal}
\affiliation{${^5}$ Perimeter Institute for Theoretical Physics, 31 Caroline Street North
Waterloo, Ontario N2L 2Y5, Canada}
\affiliation{${^6}$ Institut Lagrange de Paris (ILP), Sorbonne Universit\'es, 98 bis bd Arago, 75014 Paris, France}
\affiliation{${^7}$ Dipartimento di Fisica, ``Sapienza'' Universit\`a di Roma \& Sezione INFN Roma1, Piazzale Aldo Moro 5, 00185, Roma, Italy}

\email{richard.brito@aei.mpg.de}

\begin{abstract}
  Ultralight bosons can induce superradiant instabilities in spinning
  black holes, tapping their rotational energy to trigger the growth
  of a bosonic condensate. Possible observational imprints of these
  boson clouds include (i) direct detection of the nearly
  monochromatic (resolvable or stochastic) gravitational waves emitted
  by the condensate, and (ii) statistically significant evidence for
  the formation of ``holes'' at large spins in the spin versus mass
  plane (sometimes also referred to as ``Regge plane'') of astrophysical black
  holes. In this work, we focus on the prospects of LISA and LIGO 
  detecting or constraining scalars with mass in the range
  $m_s\in [10^{-19},\,10^{-15}]$~eV and
  $m_s\in [10^{-14},\,10^{-11}]$~eV, respectively. Using
  astrophysical models of black-hole populations calibrated to observations and black-hole
  perturbation theory calculations of the gravitational emission, we find that, in optimistic scenarios, LIGO
  could observe a stochastic background of gravitational radiation in
  the range $m_s\in [2\times 10^{-13}, 10^{-12}]$~eV, and up to $10^4$
  resolvable events in a $4$-year search if
  $m_s\sim 3\times 10^{-13}\,{\rm eV}$.  LISA could observe a
  stochastic background for boson masses in the range
  $m_s\in [5\times 10^{-19}, 5\times 10^{-16}]$, and up to $\sim 10^3$
  resolvable events in a $4$-year search if
  $m_s\sim 10^{-17}\,{\rm eV}$. LISA could further measure spins for
  black-hole binaries with component masses in the range
  $[10^3, 10^7]~M_\odot$, which is not probed by traditional 
  spin-measurement techniques.  A statistical analysis of the spin
  distribution of these binaries could either rule out scalar fields
  in the mass range $\sim [4 \times 10^{-18}, 10^{-14}]$~eV, or measure
  $m_s$ with ten percent accuracy if light scalars in the mass range
  $\sim [10^{-17}, 10^{-13}]$~eV exist.
\end{abstract}

\maketitle

\section{Introduction}

The first gravitational wave (GW) detections by the Laser
Interferometric Gravitational-wave Observatory (LIGO) are a historical
landmark.  GW150914~\cite{Abbott:2016blz},
GW151226~\cite{Abbott:2016nmj}, GW170104~\cite{Abbott:2017vtc} and the
LVT151012 trigger~\cite{TheLIGOScientific:2016pea} provided the
strongest evidence to date that stellar-mass black holes (BHs) exist
and
merge~\cite{Berti:2005ys,Cardoso:2017cfl,Cardoso:2016rao,Yunes:2016jcc,Chirenti:2016hzd}.
In this work we discuss the exciting possibility that LIGO and
space-based detectors like
LISA~\cite{AmaroSeoane:2012km,Audley:2017drz} could revolutionize our
understanding of dark matter and of fundamental interactions in the
Universe.

Ultralight bosons -- such as dark photons, the QCD axion or the
axion-like particles predicted by the string axiverse scenario --
could be a significant component of dark
matter~\cite{Arvanitaki:2009fg,Essig:2013lka,Marsh:2015xka,Hui:2016ltb}.
These fields interact very feebly with Standard Model particles, but
the equivalence principle imposes some universality in the way that
they gravitate. Light bosonic fields around spinning black holes
trigger superradiant instabilities, which can be strong enough to have
astrophysical implications~\cite{Brito:2015oca}.
Therefore, GW detectors can either probe the existence of new
particles beyond the Standard Model or -- in the absence of detections
-- impose strong constraints on their masses and
couplings~\cite{Arvanitaki:2010sy,Brito:2014wla,Arvanitaki:2014wva,Arvanitaki:2016qwi,Baryakhtar:2017ngi,Dev:2016hxv}.

Superradiance by rotating BHs was first demonstrated with a thought
experiment involving {\it
  particles}~\cite{Penrose:1969pc,Brito:2015oca}. Penrose imagined a
particle falling into a BH and splitting into two particles. If the
splitting occurs in the ergoregion, one of the fragmentation products
can be in a negative-energy state as seen by an observer at infinity,
and therefore the other fragmentation product can escape to infinity
with energy larger than the original particle.
The corresponding process involving {\it waves} amplifies any bosonic
wave whose frequency $\omega$ satisfies $0<\omega<m\Omega_{\rm H}$, where $m$
is the azimuthal index of the (spheroidal) harmonics used to separate
the angular dependence, and $\Omega_{\rm H}$ is the horizon angular
velocity~\cite{zeldovich1,zeldovich2,Brito:2015oca}.  The wave is
amplified at the expense of the BH's rotational energy.  If the wave
is trapped -- for example, through a confining mechanism like a mirror
placed at some finite distance -- the amplification process will
repeat, destabilizing the system. This creates a ``BH
bomb''~\cite{Press:1972zz,Cardoso:2004nk}.  Massive fields are
naturally trapped by their own mass, leading to a superradiant
instability of the Kerr geometry. The time scales and evolution of BH
superradiant instabilities were extensively studied by several authors
for massive
spin-0~\cite{Detweiler:1980uk,Zouros:1979iw,Cardoso:2005vk,Dolan:2007mj},
spin-1~\cite{Pani:2012vp,Pani:2012bp,East:2017ovw,East:2017mrj,Baryakhtar:2017ngi}
and spin-2 fields~\cite{Brito:2013wya}, using both analytic and
numerical methods.

\begin{figure*}[th]
\begin{center}
\begin{tabular}{c}
\epsfig{file=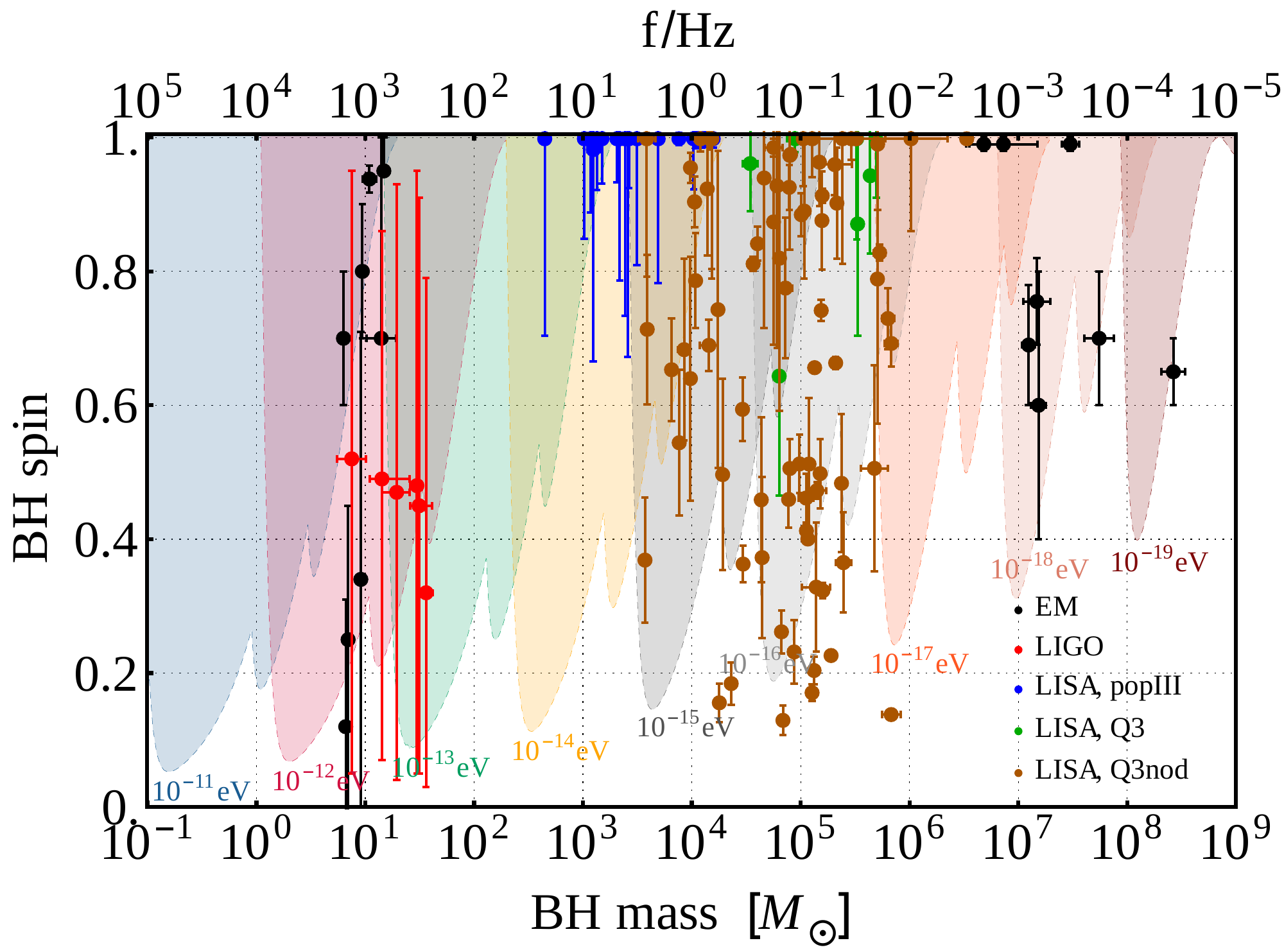,width=0.75\textwidth,angle=0,clip=true}
\end{tabular}
\caption{Exclusion regions in the BH mass-spin plane (Regge plane) for
  a massive scalar field. For each mass $m_s$, the instability
  threshold is obtained by setting the superradiant instability time
  scales for $l=m=1,\,2,\,3$ equal to a typical accretion time scale, taken to be
  $\tau=50\,{\rm Myr}$ (see main text for details). 
  Black data points (with error
  bars) are spin estimates of stellar and massive BHs obtained through
  the K$\alpha$ or continuum fitting
  methods~\cite{Brenneman:2011wz,Middleton:2015osa}. Red data points are
  GW measurements of the primary and secondary BHs from the three LIGO detections (GW150914, GW151226
  and GW170104~\cite{TheLIGOScientific:2016pea,Abbott:2017vtc}). Blue,
  green and brown data points are projected LISA measurements under the
  assumption that there are no light bosons for three different
  astrophysical black hole population models (popIII, Q3 and Q3-nod
  from~\cite{Klein:2015hvg}), as discussed in the text. We
  assume a LISA observation time $T_{\rm obs}=1\,{\rm yr}$, and to avoid cluttering we only show events for which
  LISA spin measurement errors are relatively small ($\Delta
  \chi/\chi\leq 2/3$). The top horizontal line is a frequency scale corresponding to the BH mass, $f\approx\mu/\pi$ with $\mu\sim 0.2/M$ as a reference value.
  \label{fig:Regge}}
\end{center}
\end{figure*}

For a bosonic field with mass $m_s$, superradiant instabilities are
strongest when the Compton wavelength of the massive boson
$\hbar/(m_s c)$ is comparable to the Schwarzschild radius $R=2GM/c^2$,
where $M$ is the BH mass.
Under these conditions the bosonic field can bind to the BH, forming a
``gravitational atom.''  
Instabilities can produce holes in the BH mass/spin plane (sometimes also called the BH
``Regge plane''): for a given boson mass, spinning BHs should not
exist when the dimensionless spin $\chi\equiv a/M$ is above an
instability window centered around values of order unity of the
dimensionless quantity~\cite{Brito:2015oca,Arvanitaki:2010sy}
\be
\frac{2GMm_s}{c\hbar}=1.5\frac{M}{10^6M_{\odot}}\frac{m_sc^2}{10^{-16}{\rm eV}}\,.
\ee

Typical instability windows for selected values of $m_s$ are shown as
shaded areas in Fig.~\ref{fig:Regge},
which shows the spin versus mass plane. These instability windows are obtained by requiring
that the instability acts on timescales shorter than known
astrophysical processes such as accretion, i.e. we require that the superradiant
instability time scales for scalar field perturbations with
$l=m=1,\,2,\,3$
are shorter than a typical accretion time scale, here conservatively assumed to be
the Salpeter time scale
defined below for a typical efficiency $\eta=0.1$ and Eddington rate $f_{\rm Edd}=1$ [cf.\ Eq.~\eqref{salpeter}].

In Fig.~\ref{fig:Regge}, black data points denote electromagnetic
estimates of stellar or massive BH spins obtained using either the
K$\alpha$ iron line or the continuum fitting
method~\cite{Brenneman:2011wz,Middleton:2015osa}.
Roughly speaking, massive BH spin measurements probe the
existence of instability windows in the mass range
$m_s\sim 10^{-19}$--$10^{-17}$~eV. For stellar-mass BHs, the relevant
mass range is $m_s\sim 10^{-12}$--$10^{-11}$~eV.
Red data points are LIGO 90\% confidence levels for the spins of the
primary and secondary BHs in the three merger events detected
so far (GW150914, GW151226 and
GW170104~\cite{TheLIGOScientific:2016pea,Abbott:2017vtc}). 
For LIGO BH binaries accretion should not be important. In such case, our choice for the reference timescale $t_S$ is conservative: more accurate and stringent constraints can be imposed by comparing the instability timescale with the Hubble time or with the age of the BHs.
On the other hand, in Fig.~\ref{fig:Regge} we do not include the remnant BHs detected by LIGO because the
observation time scale of the latter is obviously much shorter than the superradiant instability time scale, and
therefore post-merger observations can not be used to place constraints on
 the boson mass.

Blue, green and brown data points are projected LISA measurements
for three different astrophysical black-hole population models
(popIII, Q3, Q3-nod) from~\cite{Klein:2015hvg}, assuming one year
of observation.
The main point of Fig.~\ref{fig:Regge} is to highlight one of the most
remarkable results of this work: LISA BH spin measurements cover the
intermediate mass range (roughly $m_s\sim 10^{-13}$--$10^{-16}$~eV,
with the lower and upper bounds depending on the astrophysical model,
and more specifically on the mass of BH seeds in the early Universe),
unaccessible to electromagnetic observations of stellar and
massive BHs. In other words, LISA's capability to measure the
mass and spin of binary BH components out to cosmological
distances\footnote{We do not study holes in the Regge plane for LIGO
  because spin magnitude measurements for the binary
  components are expected to be poor, even with third-generation
  detectors~\cite{Vitale:2016avz,Vitale:2016icu}, and they overlap in mass with
  existing EM spin estimates.} implies that LISA can also probe the existence of
light bosonic particles in a large mass range that is not
accessible by other BH-spin measurement methods. In
Sec.~\ref{sec:holes} below we quantify this expectation with a more
detailed Bayesian model-selection analysis, showing in addition that
(if light bosons exist) LISA could measure their mass with $\sim 10$\%
accuracy.

We note that electromagnetic measurements of black-hole spins also provide constraints on the scalar field masses that partly overlap with constraints derived in this paper. For example, the spin measurements of stellar mass BHs disfavor the existence of a scalar field with masses between roughly $2\times 10^{-11}\,{\rm eV}>m_s>6\times 10^{-13}\,{\rm eV}$~\cite{Arvanitaki:2014wva} and  $4\times 10^{-17}\,{\rm eV}>m_s>5\times 10^{-20}\,{\rm eV}$ for massive BHs. However, GW spin measurements and constraints rely on fewer astrophysical assumptions (e.g. on the accretion disk and its spectrum) than electromagnetic constraints, and are therefore more robust. On the other hand, while electromagnetic observations of stellar mass BHs partly overlap with the GW constraints from LIGO, the electromagnetic observation of massive BHs probe lower scalar field masses that the ones coming from GW observations, and are thus complementary to the constraints that we estimate in this paper. Fig.~\ref{fig:Regge} shows that electromagnetic and GW observations should be considered jointly to build evidence for or against the existence of a scalar field with a given mass.

An even more exciting prospect is the {\em direct detection} of the GWs produced by a
BH-boson condensate system~\cite{Arvanitaki:2014wva,Arvanitaki:2016qwi,Baryakhtar:2017ngi}. Through superradiance, energy and angular
momentum are extracted from a rotating BH and the number of bosons
grows exponentially, producing a bosonic ``cloud'' at distance
$\sim \hbar^2(2GMm_s^2)^{-1}$ from the BH.  This non-axisymmetric cloud
creates a time-varying quadrupole moment, leading to long-lasting,
monochromatic GWs with frequency determined by the boson mass. Thus, the existence of
light bosons can be tested (or constrained) \emph{directly} with GW
detectors.  

To estimate the detectability of these signals we need careful
estimates of the signal strength and astrophysical models for
stellar-mass and massive BH populations.
Here we compute the GW signal produced by superradiant instabilities
using GW emission models in BH perturbation
theory~\cite{Yoshino:2013ofa}, which are expected to provide an
excellent approximation for all situations of physical
interest~\cite{Brito:2014wla,East:2017ovw,East:2017mrj}. On the
astrophysical side, we adopt the same BH formation
models~\cite{Barausse:2012fy} that were used in
previous LISA studies~\cite{Klein:2015hvg,Tamanini:2016,Berti:2016lat,Babak:2017tow,Gair:2017ynp}.
As shown below, semicoherent searches with LISA (LIGO) could detect
individual signals at luminosity distances as large as $\sim 2$~Gpc
($\sim 200$~Mpc) for a boson of mass $10^{-17} (10^{-13})$~eV
(compare this with the farthest estimated distance for LIGO BH binary
merger detections so far, the $880^{+450}_{-390}$~Mpc of
GW170104~\cite{Abbott:2017vtc}).

The plan of the paper is as follows. In Sec.~\ref{sec:bosonGW} we
outline our calculation of gravitational radiation from bosonic
condensates around rotating BHs. In Sec.~\ref{sec:SMBHs} and
Sec.~\ref{sec:Irina} we present our astrophysical models of massive
and stellar-mass BH formation, respectively. Our predictions for rates
of boson-condensate GW events detectable by LISA and LIGO, either as
resolvable events or as a stochastic background, are given in
Sec.~\ref{sec:rates}. In Sec.~\ref{sec:holes} we use a Bayesian model
selection framework to quantify how LISA spin measurements in BH
binary mergers can either exclude certain boson mass ranges by looking
at the presence of holes in the Regge plane, or (if bosons exist in
the Universe) be used to estimate boson masses.  
We conclude by summarizing our main results and identifying some
promising avenues for future work.

In the following, we use geometrized units $G=c=1$.

\section{Gravitational waves from bosonic condensates around black holes}
\label{sec:bosonGW}

In general, the development of instabilities must be followed through
non-linear evolutions.  Numerical studies of the development of
superradiant instabilities are still in their infancy~(see
e.g.~\cite{Okawa:2014nda,Zilhao:2015tya,Sanchis-Gual:2015lje,Bosch:2016vcp,East:2017ovw,East:2017mrj}),
mainly because of the long instability growth time for scalar
perturbations, which makes simulations computationally prohibitive.  If
we restrict attention to near-vacuum environments, the scalar cloud
around the spinning BH can only grow by tapping the BH's rotational
energy. Standard arguments~\cite{Christodoulou:1972kt} imply
that the BH can lose at most $29\%$ of its mass.  For the process at hand, it turns out that the cloud can store at most $\sim 10\%$ of the BH's mass~\cite{East:2017ovw,Herdeiro:2017phl},  therefore  the
spacetime is described to a good approximation by the Kerr metric, and
perturbative calculations are expected to give good estimates of the
emitted radiation~\cite{Yoshino:2013ofa,Brito:2014wla}.
These expectations were recently validated by nonlinear numerical
evolutions in the spin-1 case~\cite{East:2017ovw,East:2017mrj}, where
the instability growth time scale is faster. Reassuringly, these
numerical simulations are consistent with qualitative and quantitative
predictions from BH perturbation
theory~\cite{Pani:2012vp,Pani:2012bp,Baryakhtar:2017ngi}.
In summary, a body of analytic and numerical work justifies the use
of calculations in BH perturbation theory to estimate the
gravitational radiation emitted by bosonic condensates around Kerr
BHs. We now turn to a detailed description of this calculation.

\subsection{Test scalar field on a Kerr background}

Neglecting possible self-interaction terms or couplings to other
fields, the action describing a \emph{real} scalar field minimally
coupled to gravity is
\be
S=\int d^4x \sqrt{-g} \left( \frac{R}{16\pi}-\frac{1}{2}g^{\mu\nu}\Psi^{}_{,\mu}\Psi^{}_{,\nu} - \frac{\mu^2}{2}\Psi^2\right)\,.\label{eq:MFaction}
\ee
Here we defined a parameter
\be
\mu=m_s/\hbar\,,
\ee
which has dimensions of an inverse mass (in our geometrized units) .  The field equations derived from
this action are $\nabla_{\mu}\nabla^{\mu}\Psi =\mu^2\Psi$ and
$G^{\mu\nu}=8\pi T^{\mu\nu}$, with
\begin{equation}
T^{\mu \nu}=\Psi^{,\mu}\Psi^{,\nu}-\frac{1}{2}g^{\mu\nu}\left( \Psi_{,\alpha}\Psi^{,\alpha}+{{\mu}^2}\Psi^{2}\right)\,. \label{Tmunu}
\end{equation}

In the test-field approximation, where the scalar field propagates on
a fixed Kerr background with mass $M$ and spin $J=aM$,
the general solution of the Klein-Gordon equation can be written as
\be\label{solKG}
\Psi=\Re\left[\int d\omega e^{-i\omega t+im\varphi}{_0}S_{\ell m \omega}(\vartheta)\psi_{\ell m \omega}(r)\right]\,,
\ee
where a sum over harmonic indices $(\ell,\,m)$ is implicit, and
$_{s}Y_{\ell m \omega}(\vartheta,\varphi)={}_{s}S_{\ell m
  \omega}(\vartheta)e^{im\varphi}$ are the spin-weighted spheroidal
harmonics of spin weight $s$, which reduce to the scalar spheroidal
harmonics for $s=0$~\cite{Berti:2005gp}.  The radial and angular
functions satisfy the following coupled system of differential
equations:
\beq
{\cal D}_\vartheta[{_0}S]
&+&\left[a^2(\omega^2-\mu^2)\cos^2\vartheta
-\frac{m^2}{\sin^2\vartheta}+\lambda\right]{_0}S=0\,,\nn\\
{\cal D}_r[\psi]&+&\left[\omega^2(r^2+a^2)^2-4aMrm\omega+a^2m^2\right.\nn\\
&&\left.-\Delta(\mu^2r^2+a^2\omega^2+\lambda)\right]\psi=0\,,\nn
\eeq
where for simplicity we omit the $(\ell,\,m)$ subscripts,
$r_\pm=M\pm \sqrt{M^2-a^2}$ denotes the coordinate location of the
inner and outer horizons, $\Delta=(r-r_+)(r-r_-)$,
${\cal D}_r=\Delta\partial_r\left(\Delta\partial_r\right)$, and
${\cal
  D}_\vartheta=(\sin\vartheta)^{-1}\partial_\vartheta\left(\sin\vartheta\partial_\vartheta\right)$.
For $a=0$, the angular eigenfunctions ${_0}S_{\ell m}(\vartheta)$ reduce
to the usual scalar spherical harmonics with eigenvalues
$\lambda=\ell(\ell+1)$.

Imposing appropriate boundary conditions, a solution to the above
coupled system can be obtained using, e.g., a continued-fraction
method~\cite{Cardoso:2005vk,Dolan:2007mj}. Because of dissipation,
this boundary value problem is non-hermitian. The solutions are
generically described by an infinite, discrete set of complex
eigenfrequencies~\cite{Berti:2009kk}
\be\label{eigenfrequency}
\omega_{\ell mn}\equiv\omega=\omega_R+i\omega_I\,,
\ee
where $n$ is the overtone number and
$\{\omega_R,\omega_I\} \in \mathbb{R}$.  In particular, this system
admits quasi-bound state solutions which become unstable -- i.e., from
Eq.~\eqref{solKG}, have $\omega_I>0$ -- for modes satisfying the
superradiant condition $\omega_R<m \Omega_{\rm H}$, with
$\Omega_{\rm H}=a/(2Mr_{+})$~\cite{Detweiler:1980uk,Dolan:2007mj}. For
these solutions the eigenfunctions are exponentially suppressed at
spatial infinity:
\be
\psi(r)\propto \frac{r^\nu e^{-\sqrt{\mu^2-\omega^2}r}}{r} \quad {\rm as} \quad r\to \infty\,,
\ee
where $\nu=M(2\omega^2-\mu^2)/\sqrt{\mu^2-\omega^2}$.  In the
small-mass limit $M\mu \ll 1$ these solutions are well approximated by
a hydrogenic spectrum~\cite{Detweiler:1980uk,Dolan:2007mj} with
angular separation constant $\lambda\simeq \ell(\ell+1)$ and frequency
\be\label{omega}
\omega\sim \mu-\frac{\mu}{2}\left(\frac{M\mu}{\ell+n+1}\right)^2+\frac{i}{\gamma_\ell}\left(\frac{am}{M}-2\mu r_+\right)(M\mu)^{4\ell+5}\,,
\ee
where $n=0,1,2...$, and $\gamma_1=48$ for the dominant unstable $\ell=1$
mode.

\subsection{Gravitational-wave emission}\label{sec:gwemission}

For a real scalar, the condensate is a source of GWs. For a
monochromatic source with frequency $\omega_R$, one can easily see by,
plugging the solution~\eqref{solKG} into the stress-energy
tensor~\eqref{Tmunu}, that the scalar field sources GWs with frequency
$2\omega_R$.  In the fully relativistic regime, gravitational
radiation can be computed using the Teukolsky
formalism~\cite{Teukolsky:1973ha}. This calculation is described in
detail here (see also~\cite{Yoshino:2013ofa,Brito:2014wla}).

In the Teukolsky formalism, gravitational radiation is encoded in the
Newman-Penrose scalar $\psi_4$, which can be decomposed as
\be
\psi_4(t,r,\Omega)=\sum_{\ell m} \rho^{4}\int^\infty_{-\infty}d\omega\sum_{\ell m}R_{\ell m\omega}(r)~_{-2}S_{\ell m\omega}(\Omega)e^{-i\omega t}\,, \label{psi4expansion}
\ee
where $\rho=(r-i a \cos\vartheta)^{-1}$. The radial function
$R(r)$ satisfies the inhomogeneous equation
\beq\label{teu_eq}
&&\Delta^{2}\frac{d}{dr}\left(\Delta^{-1}\frac{dR}{dr}\right)+\left(\frac{K^2+4i(r-M)K}{\Delta}-8i\omega r-\lambda\right)R\nn\\
&&=T_{\ell m\omega}\,,
\eeq
where again we omit angular indices for simplicity,
$K\equiv (r^2+a^2)\omega-am$,
$\lambda\equiv A_{s\ell m}+a^2\omega^2-2am\omega$, and $A_{s\ell m}$
are the angular eigenvalues. The source term $T_{\ell m\omega}$ is
given by
\beq 
T_{\ell m\omega}&\equiv& \frac{1}{2\pi}\int\, d\Omega\, dt \,
_{-2}\bar{S}_{\ell m} {\cal T} e^{i\omega t}\,, 
\eeq
where ${\cal T}$ is related to the scalar field stress-energy tensor~\eqref{Tmunu} and can be found in~\cite{Teukolsky:1973ha,Sasaki:2003xr}.

To solve the radial equation~\eqref{teu_eq} we use a Green-function
approach. The Green function can be found by considering two linearly
independent solutions of the homogeneous Teukolsky
equation~\eqref{teu_eq}, with the following asymptotic behavior (see
e.g.~\cite{Sasaki:2003xr}):
\begin{eqnarray}
R^{H}
&\to&
\begin{cases}
\Delta^2  e^{-ikr^*}\,&
$for$\,\, r\to r_+, \\
r^3 B_{\rm  out}e^{i\omega r^*}+
r^{-1}B_{\rm in} e^{-i\omega r^*}\,&
$for$\,\, r\to +\infty, \\
\end{cases}
\label{Rhor} 
\\
~\nonumber\\
R^{\infty}
&\to&
\begin{cases} 
A_{\rm  out} e^{ik r^*}+ 
\Delta^2 A_{\rm in} e^{-ik r^*}\,& 
$for$\,\, r\to r_+, \\
r^3  e^{i\omega r^*}\,& 
$for$\,\, r\to +\infty, \\
\end{cases}
\label{Rinf} 
\end{eqnarray} 
where $k=\omega-m\Omega_{\rm H}$, $\{A,B\}_{\rm{in},\rm{out}}$ are constants, and the tortoise coordinate is defined as
\be
r^{*}=
r+\frac{2Mr_{+}}{r_{+}-r_{-}}\ln{\frac{r-r_{+}}{2M}}
-\frac{2Mr_{-}}{r_{+}-r_{-}}\ln{\frac{r-r_{-}}{2M}}.
\label{tortoise}
\ee
Imposing ingoing boundary conditions at the horizon and outgoing boundary conditions at infinity, one finds that the solution of Eq.~\eqref{teu_eq} is given by~\cite{Sasaki:2003xr}
\begin{equation}
R=\frac{1}{W}
 \left\{R^{\infty}\int^r_{r_+}dr' 
 \frac{R^{H}
T_{\ell m\omega}}{\Delta^{2}} 
+ R^{H}\int^\infty_{r}dr' 
\frac{R^{\infty} T_{\ell m\omega}}{\Delta^{2}}\right\}, 
\end{equation}
where the Wronskian $W=({R^{\infty}\partial_r R^H-R^{H}\partial_r R^{\infty}})/{\Delta}$
is a constant by virtue of the homogeneous Teukolsky
equation~\eqref{teu_eq}. Using Eqs.~\eqref{Rinf} and \eqref{Rhor} one finds
\begin{equation} \label{wronskian}
W=2i\omega B_{\rm{in}}\,.
\end{equation}
At infinity the solutions reads
\begin{equation}
R(r\to\infty)
 \to\frac{r^3e^{i\omega r^*}}{2i\omega 
   B_{\rm in}}
\int^{\infty}_{r_+}dr' T_{\ell m\omega}
\frac{R^{H}}{\Delta^{2}}
\equiv \tilde Z^{\infty}r^3e^{i\omega r^*}\,.
\label{eq:Rinfty}
\end{equation}
Since the frequency spectrum of the source $T_{\ell m\omega}$ is discrete with frequency $\tilde \omega=\pm 2\omega_{\ell mn}$ and $\tilde m=\pm 2 m$, where $\omega_{\ell mn}$ are the scalar field eigenfrequencies, $\tilde Z^{\infty}$ takes the form
\be
\tilde Z^{\infty}=\sum_{\ell\tilde m n} \delta(\omega-\tilde \omega) Z_{\ell \tilde m\omega}^{\infty}\,,
\ee
and at $r\to \infty$, $\psi_4$ is given by
\be
\psi_4=\frac{1}{r}\sum_{\ell\tilde m n}Z_{\ell \tilde m\tilde \omega}^{\infty}~_{-2}Y_{\ell \tilde m\tilde \omega}e^{i\tilde \omega(r^*-t)}\,. \label{psi4infinity}
\ee
At infinity the Newman-Penrose scalar can be written as
\be\label{psi4polar}
\psi_4=\frac{1}{2}\left(\ddot{h}_{+} - i \ddot{h}_\times \right)\,,
\ee
where $h_+$ and $h_\times$ are the two independent GW polarizations. The energy flux carried by these waves at infinity is given by
\be
\frac{dE}{dt d\Omega}=\frac{r^2}{16\pi}\left(\dot{h}_{+}^2+\dot{h}_{\times}^2\right)\,.
\ee
Using equations~\eqref{psi4infinity} and~\eqref{psi4polar} we get
\be\label{flux}
\frac{dE}{dt}=\sum_{\ell\tilde m n}\frac{1}{4\pi\tilde\omega^2} |Z_{\ell \tilde m\tilde\omega}^{\infty}|^2\,.
\ee
We note that $|Z_{\ell \tilde m\tilde \omega}^{\infty}| \propto M_S /M^2$, where
$M_S$ is the total mass of the scalar cloud:
\be \label{masscloud}
M_S=\int T_t^t \sqrt{-g} drd\vartheta d\varphi\,,
\ee
and $\sqrt{-g}=(r^2+a^2\cos^2\vartheta)\sin\vartheta$ is the Kerr
metric determinant. Here we neglected
the energy flux at the horizon, which in general is
subdominant~\cite{Poisson:1994yf}. In fact, we will only need to
compute radiation at the superradiant threshold, where the flux at the
horizon -- being proportional to $k=(\omega-m\Omega_{\rm H})$ --
vanishes exactly~\cite{Teukolsky:1974yv}.

\begin{figure}[t]
\begin{center}
\begin{tabular}{c}
\epsfig{file=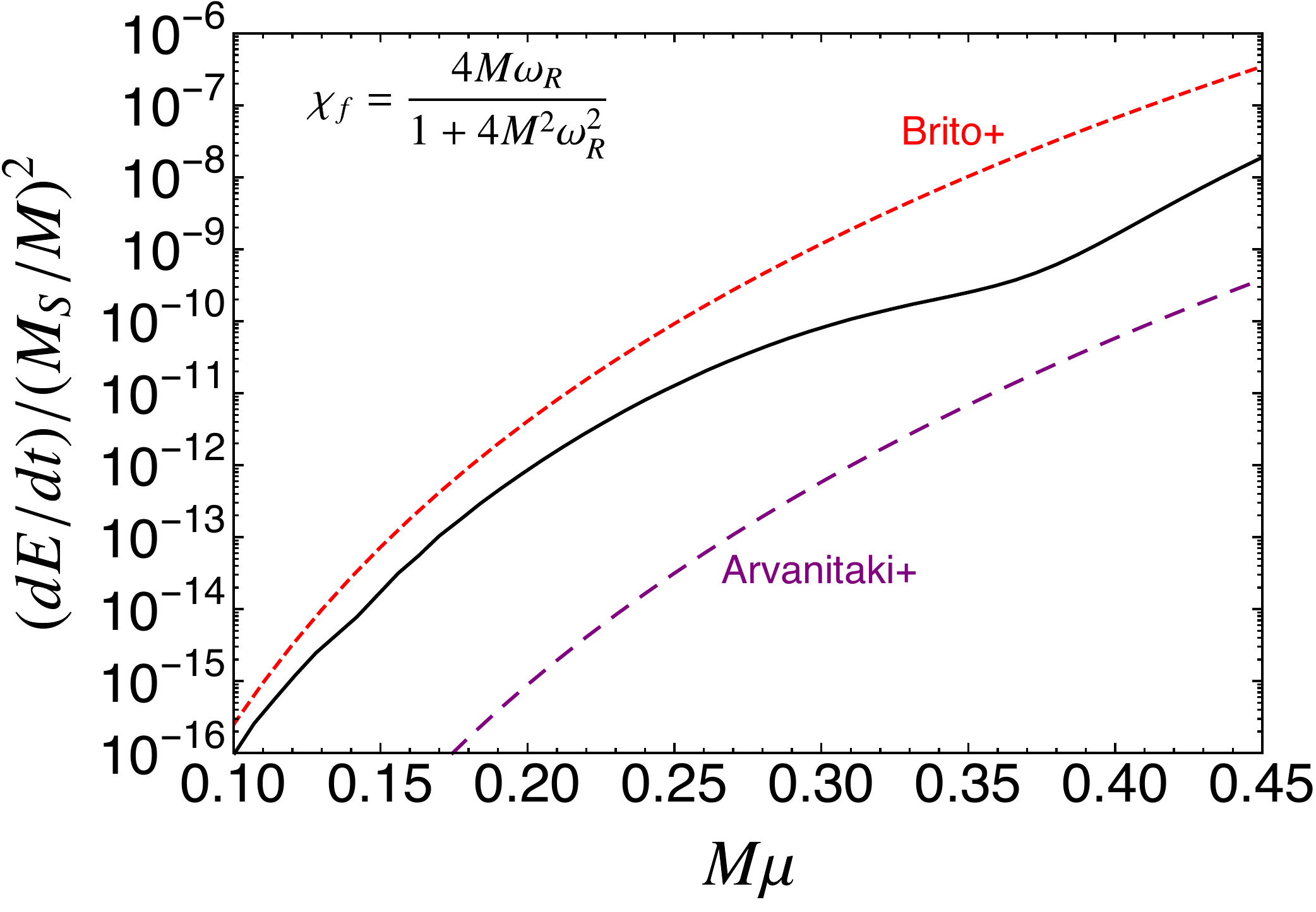,width=8cm,angle=0,clip=true}
\end{tabular}
\caption{Flux for $\ell=m=1$ and taking the first two leading order terms in the flux $\tilde{\ell}=\tilde{m}=2$ and $\tilde{\ell}=3, \tilde{m}=2$ as a
  function of the scalar mass and for the spin computed at the superradiant threshold~\eqref{finalspin}. The numerical
  results computed in this work are compared with the analytic
  formula obtained in~\cite{Brito:2014wla}, labeled ``Brito+'',
  and the one obtained in~\cite{Arvanitaki:2010sy}, labeled
  ``Arvanitaki+''.
  \label{fig:flux}}
\end{center}
\end{figure}

Figure~\ref{fig:flux} shows the dominant GW energy flux computed
numerically within the perturbative framework described above. Our
results are compared to the analytic results of
Refs.~\cite{Brito:2014wla,Arvanitaki:2010sy}. 
The flat-space approximation adopted in~\cite{Arvanitaki:2010sy}
underestimates the flux by some orders of magnitude, especially when
$\mu M\ll0.3$, for any spin. Likewise, the Schwarzschild approximation
adopted in~\cite{Brito:2014wla} overestimates the GW flux. To improve
on both approximations, in the rest of this work we will use the
numerical results, which are valid in the entire $(\chi,\mu M)$ plane
and agree with those of~\cite{Yoshino:2013ofa}.
%
\subsection{Evolution of the superradiant instability and of the BH-condensate system}\label{sec:evolution}

Current nonlinear evolutions are unable to probe the development of
the instability in the scalar case~\cite{Okawa:2014nda}. However,
since the time scales of both the superradiant instability and the GW
emission are much longer than the dynamical time scale of the BH, the
evolution of the BH-condensate system can be studied within a
quasi-adiabatic approximation~\cite{Brito:2014wla}. The scalar field
can be considered almost stationary, and its backreaction on the
geometry neglected, as long as the scalar stress-energy tensor is
small compared to the BH energy density~\cite{Brito:2014wla}.  

Recent nonlinear evolutions by East and Pretorius in the spin-1
case~\cite{East:2017ovw,East:2017mrj}, where the instability develops
more rapidly, lend support to an adiabatic treatment of the evolution
of the field. The evolution happens in two steps characterized by very
different time scales. First a scalar condensate grows around the BH
until the superradiant condition is saturated; then the condensate is
dissipated through GW emission. Neglecting accretion for simplicity,
the evolution of the system is governed by the
equations~\cite{Brito:2014wla}
\begin{eqnarray}
\left\{\begin{array}{l}
        \dot M = -\dot E_S\,, \\
	\dot M + \dot M_S = -\dot E\,, \\
	\dot J = -m\dot E_S/\omega_R\,, \\
	\dot J + \dot J_S = -m\dot E/\omega_R\,, \\
       \end{array} \right.\label{evolution}
\end{eqnarray}
where $\dot E_S=2 M_S \omega_I$ is the scalar energy flux extracted
from the horizon through superradiance. In the above equations, we
have used the fact that --~for a single $(\ell,\,m)$ mode~-- the GW
angular momentum flux is $m\dot E/\omega_R$ and that the angular
momentum flux of the scalar field extracted at the horizon is
$m\dot E_S/\omega_R$.

The system~\eqref{evolution} shows that for a superradiantly unstable
state $(\omega_I>0$) the instability will cause the BH to transfer
mass and spin to the scalar field until the system reaches the
saturation point, given by $\omega_I=0$, i.e.,
$\omega_R=m \Omega_{\rm H}$.\footnote{Fully non-linear evolutions of a
  charged scalar field around a charged BH enclosed by a reflecting
  mirror~\cite{Sanchis-Gual:2015lje,Sanchis-Gual:2016tcm,Sanchis-Gual:2016ros}
  or in anti-de Sitter spacetime~\cite{Bosch:2016vcp} have shown that
  the end-state for this system indeed consists of a scalar condensate
  around a charged BH saturating the superradiant condition. East and
  Pretorius reached the same conclusion for massive spin-1
  fields~\cite{East:2017ovw,East:2017mrj}. For complex fields, truly stationary metric solutions of the field equations describing a boson condensate saturating the superradiant condition around spinning BH have been explicitly shown to exist~\cite{Herdeiro:2014goa,Herdeiro:2015waa,Herdeiro:2016tmi}}
This process occurs on a time scale
$\tau_{\rm inst}\equiv 1/\omega_I\gg M$, and the saturation point
corresponds a final BH angular momentum
\be \label{finalspin}
J_f =  \frac{4 m M_f^3 \omega_R}{m^2+4 M_f^2\omega_R^2}< J_i\,,
\ee
where $J_{i/f}$, $M_{i/f}$ are the initial/final BH angular momentum
and mass, respectively. The system~\eqref{evolution} also shows that
the variation of the BH mass $\delta M$ is related to the variation of
the BH angular momentum $\delta J$ by
$\delta M=\frac{\omega_R}{m}\delta J$, which implies
\be \label{finalmass}
M_f=M_i-\frac{\omega_R}{m}(J_i-J_f)\,.
\ee
When the instability saturates, the total mass of the scalar cloud is
roughly given by $M_S^{\rm max}\sim M_i-M_f$, namely
\beq\label{maxmass}
M_S^{\rm max}\sim  \frac{J_i \omega_R}{m}- \frac{4M_f^3\omega_R^2}{m^2+4 
M_f^2\omega_R^2}\approx \frac{J_i \omega_R}{m}\,,
\eeq
where the last step is valid when $M_f\omega_R\ll1$. 

\begin{figure*}[htb]
\begin{center}
\begin{tabular}{c}
  \includegraphics[width=\textwidth]{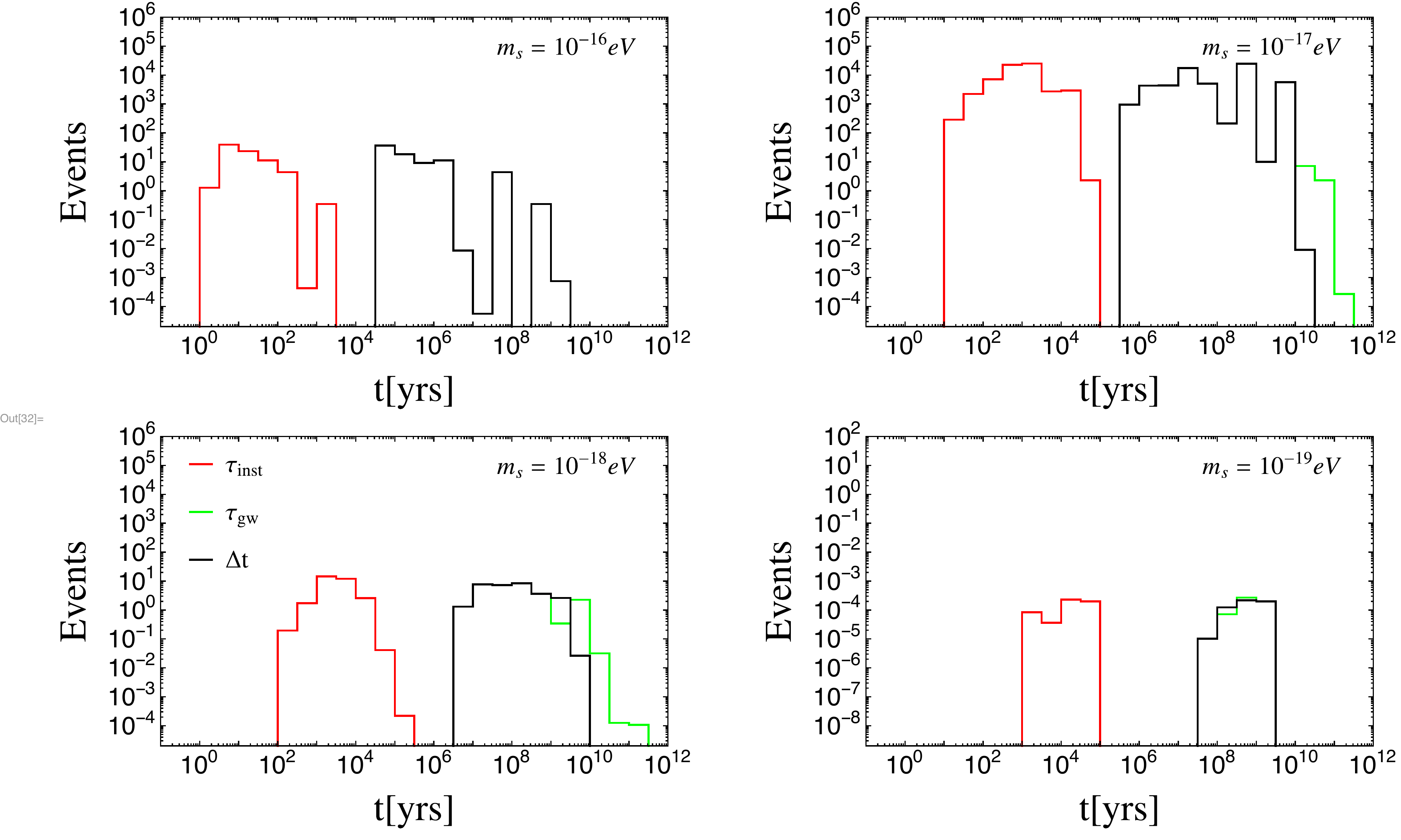}\\
\end{tabular}
\caption{Gravitational radiation time scale, instability time scale, and the
  signal duration $\Delta t$ [defined in
  Eq.~\eqref{deltat}] for detectable LISA sources and for different
  boson masses.
  \label{fig:histosd_LISA} 
}
\end{center}
\end{figure*}

After the superradiant phase, the mass and the angular momentum of the
BH remain constant [cf. Eq.~\eqref{evolution}], whereas the scalar
field is dissipated through the emission of GWs\footnote{In the
  language of~\cite{Arvanitaki:2014wva} this process corresponds to
  the ``axion+axion $\to$ graviton'' annihilation process. In our
  notation, their ``occupation number'' is $N=M_S/m_s$.}  as given by
Eq.~\eqref{flux}. We neglect GW absorption at the event horizon --
which is always sub-dominant~\cite{Poisson:1994yf} -- and GW emission
due to the transition of bosons between different energy levels, which
is also a sub-dominant effect as long as the condensate is mostly
populated by a single level~\cite{Arvanitaki:2014wva}.
By using again Eq.~\eqref{evolution}, after the superradiant phase we get
\be\label{gwflux}
\dot M_S=-\frac{d E}{dt}=-\frac{d\tilde{E}}{dt}\frac{M_S^2}{M_f^2}\,,
\ee
where we used the fact that
$|Z_{\ell m\omega}^{\infty}|^2 \propto M_S^2$ to factor out the
dependence on $M_S(t)$, and we defined
$\frac{d\tilde{E}}{dt}\equiv\frac{d E}{dt} \frac{M_f^2}{M_S^2}$. This
quantity is shown in Figure~\ref{fig:flux} and it is constant after
the superradiant phase, since it depends only on the final BH mass and
spin.
Therefore, setting $t=0$ to be the time at which the
superradiant phase saturates, the above equation yields
\begin{equation}
M_S(t)=\frac{M_S^{\rm max}}{1+t/\tau_{\rm GW}}\,, \label{Ms}
\end{equation}
where $M_S^{\rm max}$ is the mass of the condensate at the end of the
superradiant phase [cf.\ Eq.~\eqref{maxmass}] and
\begin{eqnarray}
 \tau_{\rm GW}&\approx &M_f \left(\frac{d\tilde{E}}{dt}\frac{M_S^{\rm max}}{M_f}\right)^{-1}\nn\\
 &\approx& 8\times10^5\,{\rm yr}\left[\frac{M_f}{10^6M_\odot}\right]\left[\frac{10^{-11}}{{d\tilde E}/{dt}}\right]\left[\frac{0.2M_f}{M_S^{\rm max}}\right] \label{signalduration}
\end{eqnarray}
is the gravitational radiation time scale.

\begin{figure*}[htb]
\begin{center}
\begin{tabular}{c}
  \includegraphics[width=\textwidth]{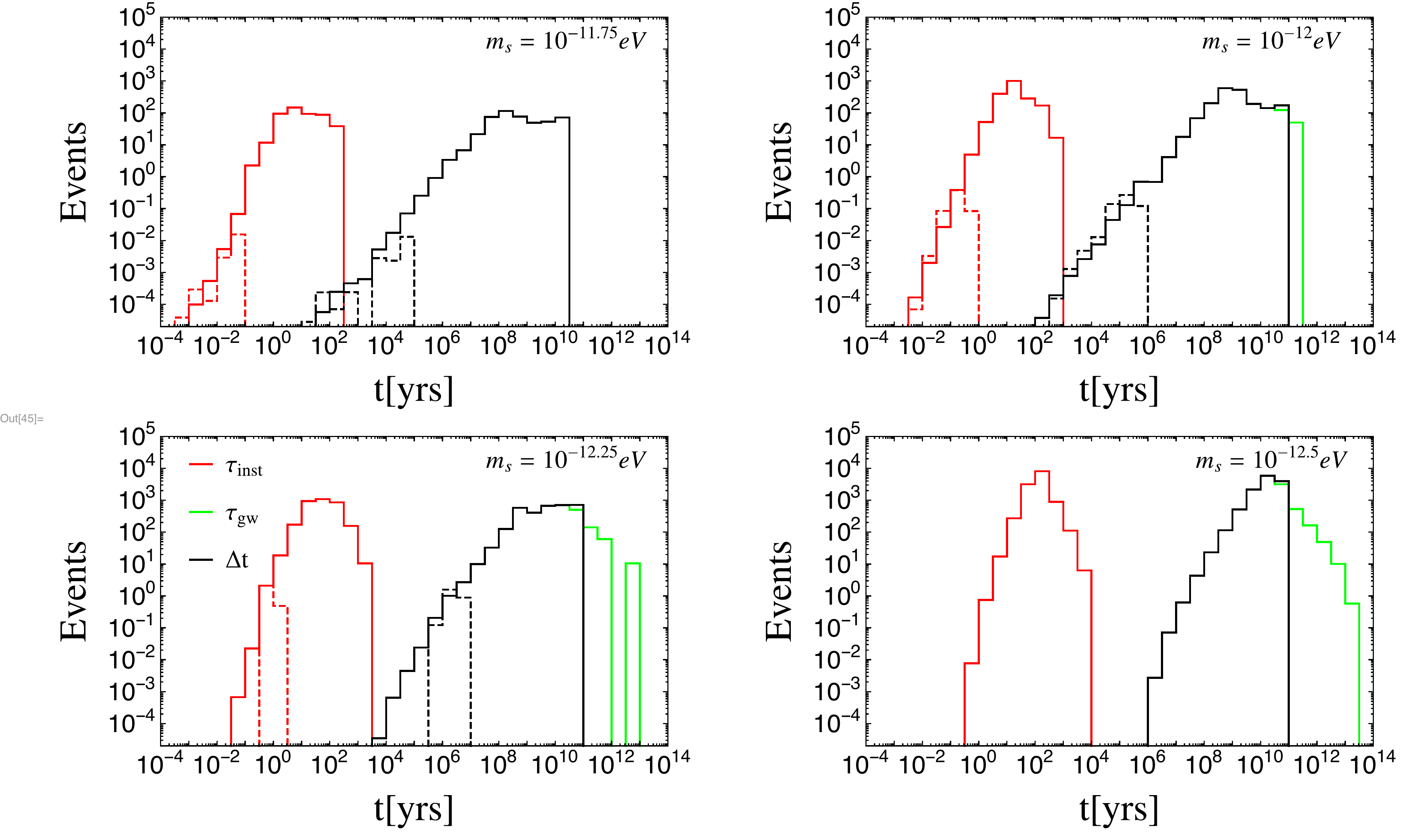}\\
\end{tabular}
\caption{Gravitational radiation time scale, instability time scale, and the
  signal duration $\Delta t$ [defined in
  Eq.~\eqref{deltat}] for detectable LIGO sources and for different boson
  masses. Dashed lines represent extragalactic sources and bold lines
  represent Galactic sources.
\label{fig:histosd_LIGO} 
}
\end{center}
\end{figure*}

Finally, we note that the self-gravity of the boson cloud will cause the GW frequency to change slightly as the cloud dissipates via GWs~\cite{Arvanitaki:2014wva,Baryakhtar:2017ngi}. The estimates of Refs.~\cite{Arvanitaki:2014wva} [see their Eq.~(28)] and~\cite{Baryakhtar:2017ngi} [see their Appendix E] suggest that, for scalar fields, this small change should not affect current continuous-wave searches. Taking these estimates and the duration of the signal of Figs.~\ref{fig:histosd_LISA} and~\ref{fig:histosd_LIGO} for resolved events, one can see that for both LIGO and LISA a vast majority of the sources will have a small positive frequency drift $\dot{f}\ll 10^{-9} \rm{Hz/s}$, which is the current upper limit on the frequency time derivative of the latest all-sky search from LIGO~\cite{Abbott:2017mnu}. However, even though this frequency drift should be very small and undetectable for most sources, the positive frequency time derivative of GWs from boson clouds could be used to distinguish them from other continuous sources, such as rotating neutron stars, which have a negative frequency drift~\cite{Palomba:2012wn}.

\subsection{Instability and gravitational radiation time scales}
\label{sec:inst_timescale}

As discussed above, the basic features of the evolution of the BH
superradiant instability in the presence of light bosons 
can be understood as a two-step process, governed by two
different time scales. The first time scale is the typical e-folding
time of the superradiant instability given by $\tau_{\rm inst}\equiv 1/\omega_I$,
where in the $M\mu\ll 1$ limit, $\omega_I$ is the imaginary part of Eq.~\eqref{omega}.  The
boson condensate grows over the time scale $\tau_{\rm inst}$ until the
superradiant condition is saturated.  Subsequently, the condensate is
dissipated through GW emission over a time scale $\tau_{\rm GW}$ given
by Eq.~\eqref{signalduration}. In the $M\mu\ll 1$ limit,
$d\tilde{E}/dt= (484 + 9 \pi^2) /23040 (\mu M)^{14} \simeq 0.025 (\mu
M)^{14}$~\cite{Yoshino:2013ofa,Brito:2014wla}.
Thus, using Eqs.~\eqref{omega}, \eqref{maxmass},
\eqref{signalduration} and reinstating physical units, the two most
relevant time scales of the system are of the order
\begin{eqnarray}
&\tau_{\rm inst} \sim 10^{5} {\rm yr}  \left( M_6^8  \mu_{17}^9  \chi\right)^{-1}\,, \label{tauSR}\\
&\tau_{\rm GW}\sim 5\times 10^{11} {\rm yr}  \left(M_6^{14} \mu_{17}^{15} \chi\right)^{-1}\,, \label{tauGW}
\end{eqnarray}
where $M_6=M/(10^6 M_\odot)$ and $\mu_{17}= m_s/(10^{-17} {\rm eV})$
and $\chi\ll 1$. 

These relations are still a reasonably good approximation when
$M\mu\sim 1$ and $\chi\sim 1$. They show that there is a clear
hierarchy of time scales ($\tau_{\rm GW}\gg\tau_{\rm inst}\gg M$), and
this is important for two reasons. First of all it is crucial that
$\tau_{\rm GW}\gg\tau_{\rm inst}$, otherwise the boson condensate
would not have time to grow.  Second, the time scale hierarchy
justifies the use of an adiabatic approximation to describe the
evolution.

Beyond the instability and gravitational radiation time scales, from
the point of view of detection it is important to estimate the
distribution of {\em signal durations} $\Delta t$. For LIGO we can
safely neglect accretion, because accreted matter is not expected to
significantly alter the birth spin of stellar-mass
BHs~\cite{King:1999aq}.  We can also neglect the effect of mergers,
since mergers affect a very small fraction of the overall population of
isolated
BHs~\cite{Dominik:2012kk,Dominik:2013tma,Dominik:2014yma,Belczynski:2015tba,Belczynski:2016obo},
and LIGO data already suggest that multiple mergers should be
unlikely~\cite{Gerosa:2017kvu,Fishbach:2017dwv}.  Therefore, for LIGO
we will simply assume $\Delta t=\min\left(\tau_{\rm GW}, t_0\right)$, where
$t_0\approx 13.8\,{\rm Gyr}$ is the age of the Universe.

For massive BHs that radiate in the LISA band, both mergers and
accretion are expected to be
important~\cite{Berti:2008af,Sesana:2014bea}. Therefore we
conservatively assume that whenever an accretion event or a merger
happens the boson-condensate signal is cut short, and for LISA we
define
\begin{equation}
 \Delta t=\left\langle\min\left(\frac{\tau_{\rm GW}}{N_m+1},t_S,
     t_0\right)\right\rangle\,, 
\label{deltat}
\end{equation}
where the signal duration $\tau_{\rm GW}$ in the absence of mergers
and accretion is given by Eq.~\eqref{signalduration},
$\langle...\rangle$ denotes an average weighted by the probability
distribution function of the Eddington ratios, $t_S$ is the
``Salpeter'' accretion time scale [Eq.~\eqref{salpeter}], and $N_m$ is
the average number of mergers expected in the interval
$[t-\tau_{\rm GW}/2,t+\tau_{\rm GW}/2]$, $t$ being the cosmic time
corresponding to the cosmological redshift $z$ of the GW source. Note that this definition also enforces the
obvious fact that the signal cannot last longer than the age of the
Universe ($\Delta t\leq t_0$).
We also note that the estimates of Refs.~\cite{Arvanitaki:2014wva,Baryakhtar:2017ngi} suggest that the close passage of a stellar-mass compact object around the massive BH could affect the boson cloud when $M\mu\ll 0.1$. This part of the parameter space is mostly irrelevant for our results, and so we neglect this contribution.
Moreover, estimates of the rates of extreme mass-ratio inspirals predict at most a few hundred such close passages per Gyr per galaxy~\cite{Babak:2017tow}. Therefore, the average timescale between these events is  $\gtrsim10^7$ yr. This is comparable with the accretion timescale [Eq.~\eqref{salpeter}], which we have already taken into account. Thus, we expect our results to be robust against inclusion of this effect.  In addition, stars and compact objects could, in principle, affect the boson cloud also at larger  orbital distances, comparable to the peak of the cloud $R\sim 4M/(M\mu)^2$~\cite{Brito:2014wla}. This could also become relevant for $M\mu\ll 0.1$, but even in this case passages of stars at $R\sim 1000M$ or larger are expected to be quite rare. Indeed, tidal disruption of stars are about $10^{-5}$ per yr per galaxy~\cite{tde}, hence stars at distances $R\sim 1000M$  from the BH should
only appear roughly every $10^5$ yr. We have checked that even if we include this effect by adding an extra timescale $\sim 10^5$ yr to Eq.~\eqref{deltat}, the background and the resolved event rates would only decrease by about an order of magnitude (and only for $M\mu\ll 0.1$), thus leaving our conclusions unchanged.

Figures~\ref{fig:histosd_LISA} and~\ref{fig:histosd_LIGO} show
histograms of $\tau_{\rm inst}$, $\tau_{\rm GW}$ and $\Delta t$ for
resolvable sources with SNR $\rho\geq 8$
[cf. Eq.~\eqref{SNR_Enrico}]. When computing the SNR, we use an
observation time $T_{\rm obs}=2$~yr for LIGO and $T_{\rm obs}=4$~yr
for LISA.
We adopt the LISA noise power spectral density specified in the ESA
proposal for L3 mission concepts~\cite{Audley:2017drz} and the design
sensitivity of Advanced LIGO~\cite{Aasi:2013wya}.
The events are binned by gravitational radiation time scale $\tau_{\rm GW}$, instability
time scale $\tau_{\rm inst}$, and signal duration $\Delta t$,
as defined in Eq.~(\ref{deltat}).
For concreteness, in the plot we focus on the most optimistic
astrophysical model, and we neglect the confusion noise due to the
stochastic background produced by these sources [cf.~\cite{Brito:2017wnc}].
For LIGO we show both Galactic and extragalactic sources.

The signal duration $\Delta t$ is typically equal to the
gravitational radiation time scale $\tau_{\rm GW}$, and (as
anticipated) much longer than the instability time scale
$\tau_{\rm inst}$. Since for LIGO we neglect the effects of mergers and
accretion, the only visible difference between $\Delta t$ and
$\tau_{\rm GW}$ is due to the fact that we cut off the signal when its typical time scale is longer than the age of the Universe (i.e., as mentioned above, we set
$\Delta t=t_0$ if $\tau_{\rm GW}>t_0$).
For LISA there are more subtle effects related to accretion and
mergers [cf. Eq.~(\ref{deltat})], but Figs.~\ref{fig:histosd_LISA}
and~\ref{fig:histosd_LIGO} demonstrate that 
the signal duration $\Delta t$ is always much longer than
the instability time scale $\tau_{\rm inst}$, as suggested by the
rough estimates of Eqs.~(\ref{tauSR}) and (\ref{tauGW}).

\subsection{Gravitational waveform}

Since the GW signal from boson condensates is quasi-monochromatic,
 we can can compute the (average)
 signal-to-noise ratio (SNR) as~\cite{Stroeer:2006rx,Ruiter:2007xx}
 \beq\label{SNR_Enrico} {\rho}\simeq\left\langle\frac{h\sqrt{T_{\rm
         overlap}}}{\sqrt{S_h(f)}}\right\rangle\,, \eeq where $h$ is the
 root-mean-square (rms) strain amplitude; $S_h(f)$ is the noise power
 spectral density at the (detector-frame) frequency $f$ of the signal,
 which is related to the source-frame frequency
 $f_s\equiv {\omega}/(2\pi)$ by $f=f_s/(1+z)$ ($z$ being the
 redshift); $T_{\rm overlap}$ is the overlap time between the observation period $T_{\rm obs}$ and the signal
 duration $\Delta t(1+z)$ [in the detector frame, hence the factor $1+z$ multiplying the
 signal duration $\Delta t$ in the source frame]; and
 $\langle\dots\rangle$ denotes an average over the possible overlap
 times. In practice, when our astrophysical models predict that a signal
 should overlap with the observation window, we compute this average by randomizing the signal's
 starting time with uniform probability distribution in the interval
 $[-\Delta t(1+z), T_{\rm obs}]$ (where we assume, without loss of generality, that $t=0$
 is the starting time of the observation period).

 Coherent searches for almost-monochromatic sources are
 computationally expensive, and normally only feasible when the intrinsic parameters of the source and its
sky location are known. For {\em all-sky} searches, where the properties and location of
 the sources are typically unknown, it is more common to use
 semicoherent methods, where the signal is divided in ${\cal N}$
 coherent segments with time length $T_{\rm coh}$. The typical
 sensitivity threshold, for signals of duration $\Delta t(1+z) \gg T_{\rm obs}$, is [cf.  e.g.~\cite{Palomba:2012wn}]
\be\label{hthr}
h_{\rm thr}\simeq \frac{25}{{\cal N}^{1/4}} \sqrt{\frac{S_h(f)}{T_{\rm coh}}}\,,
\ee
where $h_{\rm thr}$ is the minimum rms strain amplitude detectable
over the observation time ${\cal N}\times T_{\rm coh}$. This criterion
was used, for example, in~\cite{Arvanitaki:2014wva}.
In the following we consider both cases (a full coherent search and a
semicoherent method) in order to bracket uncertainties due to specific
data analysis choices. For the semicoherent searches we only consider
events for which  $\Delta t (1+z) \gg T_{\rm obs}$ [since the threshold gived by eq.~\eqref{hthr}
only holds for long-lived signals].

A useful quantity to compare the sensitivity of different searches
independently of the data-analysis technique and the quality and
amount of data is the so-called ``sensitivity depth,'' defined
by~\cite{Behnke:2014tma}
\be
\mathcal{D}(f)=\frac{\sqrt{S_h (f)}}{h_{\rm thr}}\,.
\label{SenDep}
\ee
For example, the average sensitivity depth of the last {\sc Einstein@Home}
search was
$\mathcal{D}\approx 35
\rm{Hz}^{-1/2}$~\cite{TheLIGOScientific:2016uns}.
 
To compute $h$, we first use
Eqs.~\eqref{psi4expansion},~\eqref{psi4infinity} and \eqref{psi4polar}
to get a combination of the two GW polarizations,
\be
H\equiv h_{+}-ih_{\times}=-\frac{2}{\tilde\omega^2r}\sum_{\ell\tilde m n}Z_{\ell \tilde m\tilde \omega}^{\infty}~_{-2}Y_{\ell \tilde m\tilde \omega}e^{i\tilde\omega(r^*-t)}\,.
\ee
In the following we will omit the sum over $\ell\tilde m n$ for ease of notation.
Let us focus on a single scalar field mode\footnote{In this work we will focus on the mode with the smallest instability time scale $\ell=m=1$, which should be the dominant source of GW radiation~\cite{Arvanitaki:2014wva}.}. If the scalar field has
azimuthal number $m$ and real frequency $\omega_R$, the GW emitted by
the scalar cloud will have azimuthal number $\tilde m=\pm 2 m$ and
frequency $\tilde \omega=\pm 2 \omega_R$. Defining $Z^{\infty}=|Z|e^{-i\phi}$, where $|Z|$ and $\phi$ are
both real, we have
\begin{eqnarray}
H=-\frac{2|Z|}{\tilde\omega^2r}&&\left({}_{-2}Y_{\ell\tilde m\tilde \omega}e^{i\left[\tilde \omega(r^*-t)+\phi\right]}\right.\nn\\
&&\left.+_{-2}Y_{\ell-\tilde m-\tilde \omega}e^{-i\left[\tilde \omega(r^*-t)+\phi\right]}\right)\,,
\end{eqnarray}
where we used the fact that
$Z_{\ell -\tilde m-\tilde \omega}^{\infty}={Z}_{\ell \tilde m\tilde
  \omega}^{\infty}$. Since
$_{s}Y_{\ell \tilde m \tilde\omega}(\vartheta,\varphi)={}_{s}S_{\ell \tilde m
  \tilde \omega}(\vartheta)e^{im\varphi}$ and $S$ is a real function for real
$\tilde\omega$, we get
\beq
h_{+}=\Re(H)\equiv&-&\frac{2|Z|}{\tilde\omega^2r}\left({}_{-2}S_{\ell
    \tilde{m}\tilde{\omega}}+{}_{-2}S_{\ell
    -\tilde{m}-\tilde{\omega}}\right)\nn\\
&\times&\cos\left[\tilde\omega(r^*-t)+\phi+\tilde m\varphi\right]\,,\\
h_{\times}=\Im(H)\nn
\equiv&-&\frac{2|Z|}{\tilde\omega^2r}\left({}_{-2}S_{\ell
    \tilde{m}\tilde{\omega}}-{}_{-2}S_{\ell
    -\tilde{m}-\tilde{\omega}}\right)\nn\\
&\times&\sin\left[\tilde\omega(r^*-t)+\phi+\tilde m\varphi\right]\,.
\eeq
The GW strain measured at the detector is
\be
h=h_+F_++h_{\times}F_{\times}\,,
\ee
where $F_{+,\times}$ are pattern functions that depend on the orientation of the detector and the direction of the source. To get the rms strain of the signal we angle-average over source and detector directions and use $\left<F_+^2\right>=\left<F_{\times}^2\right>=1/5$, $\left<F_+F_{\times}\right>=0$, $\left<|_sS_{\ell \tilde m\tilde\omega}|^2\right>=1/(4\pi)$ and $\left<\cos^2\left[\tilde\omega(r^*-t)+\phi+\tilde m\varphi\right]\right>=\left<\sin^2\left[\tilde\omega(r^*-t)+\phi+\tilde m\varphi\right]\right>=1/2$.
We then obtain
\begin{align}\label{waveform}
&h\simeq\left<h^2\right>^{1/2}=\left(\frac{2|Z|^2}{5\pi\tilde\omega^4r^2}\right)^{1/2}=\left(\frac{4\dot{E}}{5\tilde\omega^2r^2}\right)^{1/2}\,,
\end{align}
where $\dot{E}$ is given in Eq.~\eqref{flux}, which for a single scalar mode reads $\dot{E}=\sum_{\ell}|Z_{\ell}|^2/(2\pi\tilde \omega^2)$. 
Finally, let us factor out the BH mass and the mass of the scalar
condensate: $|Z|=A(\chi,\mu M) (M\tilde\omega)^2 M_S/M^2$, where $A(\chi,\mu M)$ is a dimensionless
quantity. The final expression for the rms strain reads
\begin{equation}
h=\sqrt{\frac{2}{5\pi}}
\frac{M}{r} \frac{M_S}{M}
A(\chi,\mu M)\,. \label{strain}
\end{equation}
We conservatively assume that the GWs observed at the detector are
entirely produced after the saturation phase of the
instability. Therefore, we compute $h$ using the \emph{final} BH mass
and spin, as computed in Eqs.~\eqref{finalmass} and~\eqref{finalspin},
respectively.  Larger initial spins imply that a larger fraction of the BH mass is
transferred to the scalar condensate [cf.\ Eq.~\eqref{maxmass}]. So,
for a given scalar field mass and initial BH mass, the strain grows
with the initial spin.

Equation~\eqref{strain} is valid for any interferometric detector for
which the arms form a 90-degree angle, such as Advanced LIGO. For a
triangular LISA-like detector the arms form a 60-degree angle, and we
must multiply all amplitudes by a geometrical correction factor
$\sqrt{3}/2$~\cite{Cutler:1997ta,Berti:2005ys}.  Additionally, since
we sky-average the signal, we will use an effective non-sky-averaged
noise power spectral density, obtained by multiplying LISA's sky-averaged $S_h$ by
$3/20$~\cite{Berti:2004bd}.
The analysis presented below takes into account these corrective factors.

\begin{figure*}[htb]
\begin{center}
\begin{tabular}{c}
\epsfig{file=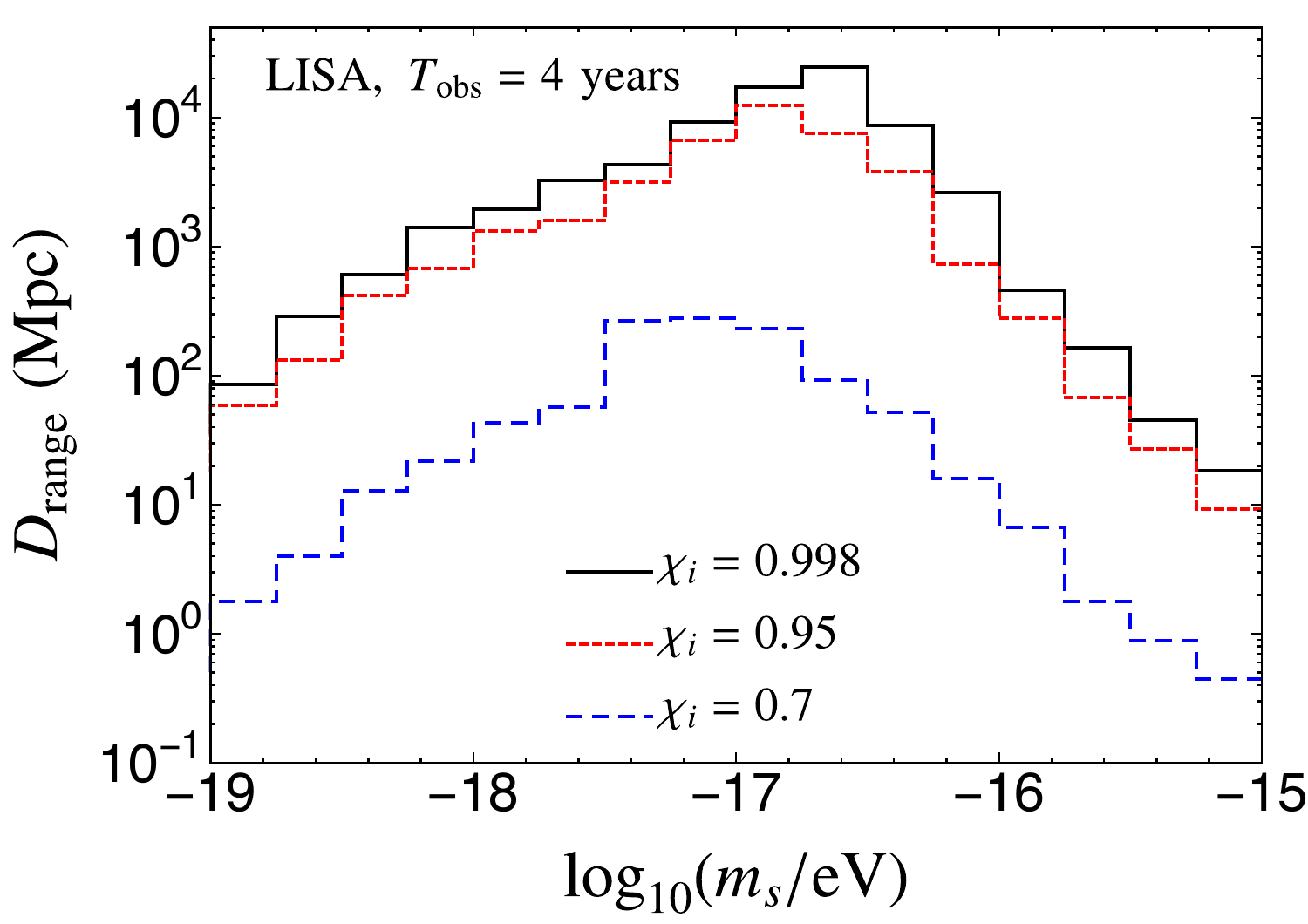,width=\columnwidth,angle=0,clip=true}
\epsfig{file=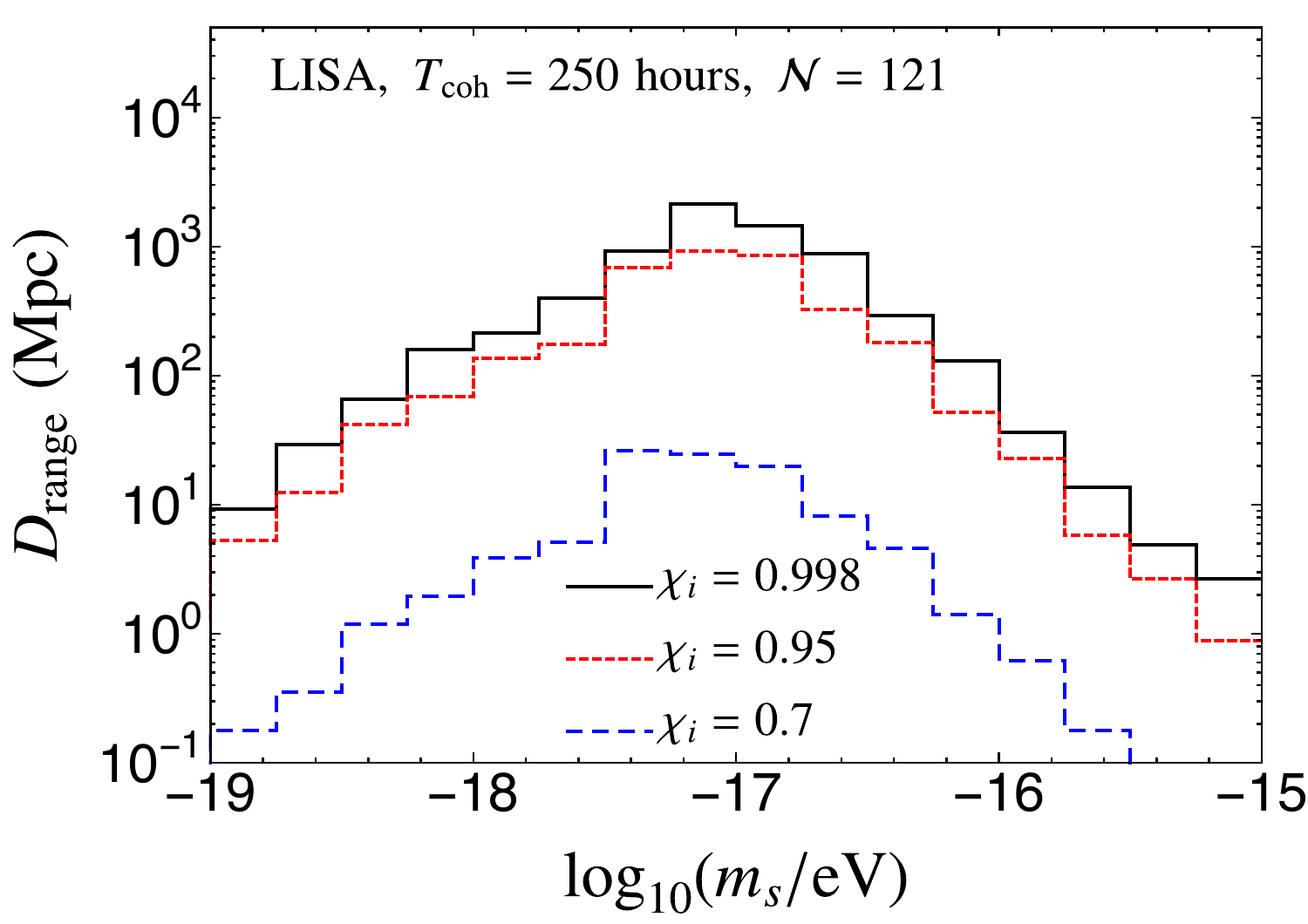,width=\columnwidth,angle=0,clip=true}\\
\epsfig{file=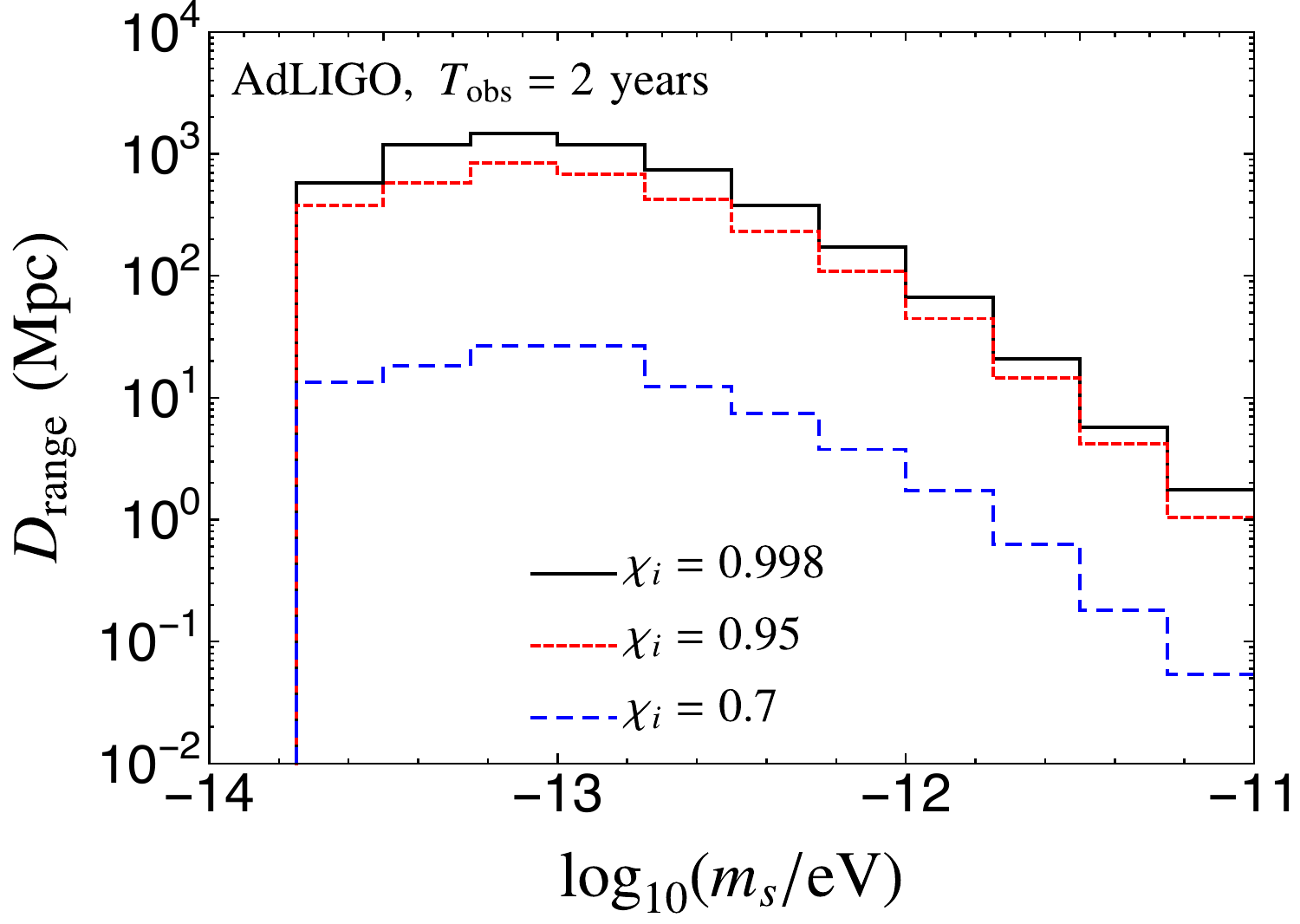,width=\columnwidth,angle=0,clip=true}
\epsfig{file=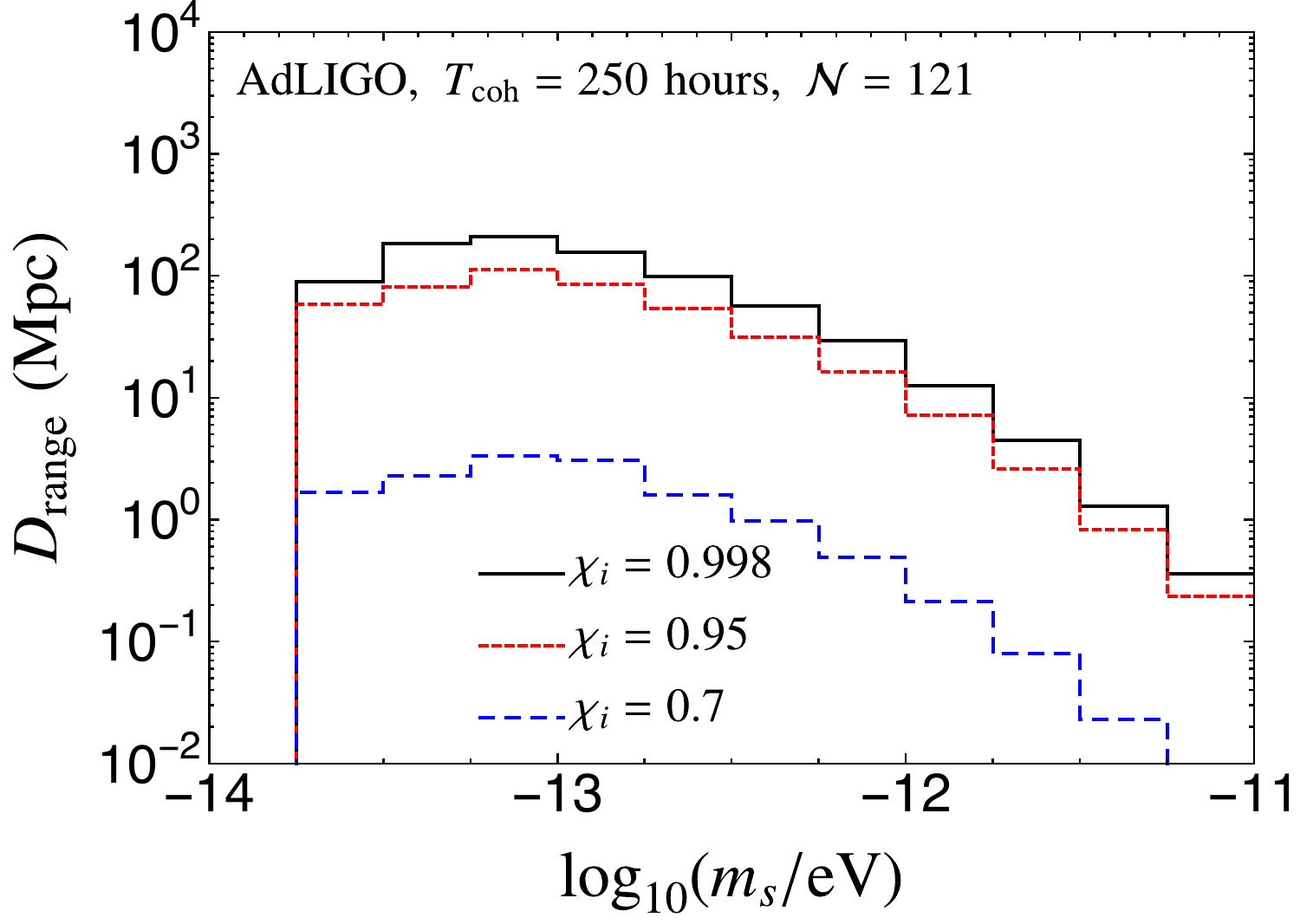,width=\columnwidth,angle=0,clip=true}
\end{tabular}
\caption{Angle-averaged range $D_{\rm range}$ for LISA (top) and
  Advanced LIGO at design sensitivity (bottom) computed for selected
  initial BH spin ($\chi_i=0.998,\,0.95,\,0.7$). Left
  panels: the range is computed using a coherent search over an
  observation time $T_{\rm obs}=4$ yr (for LISA) and $T_{\rm obs}=2$
  yr (for LIGO). Right panels: we assume a semicoherent search with
  ${\cal N}=121$ coherent segments of duration $T_{\rm
    coh}=250\,{\rm hr}$.
  \label{reachradius}}
\end{center}
\end{figure*}
%
 
\subsection{Cosmological effects}

Since some sources can be located at non-negligible redshifts,
the root-mean-square strain amplitude of Eqs.~\eqref{waveform}
and~\eqref{strain} must be corrected to take into account cosmological
effects, which affect the propagation of the waves to the
detector~\cite{Maggiore:1900zz}. These effects have two main
consequences.

First, the frequency $f$ of the signal as measured at the detector's
location (``detector frame'') is redshifted with respect to the
emission frequency $f_s$ in the ``source-frame'', i.e. $f=f_s/(1+z)$.

Second, in the strain amplitude given by Eq.~\eqref{strain}, the
distance $r$ to the detector should be interpreted as the comoving
distance, which for a flat Friedmann-Lemaitre-Robertson-Walker model
is given by
\be D_c(z)=D_H\int_0^z
\frac{dz'}{\sqrt{\Delta(z')}}, 
\ee 
where $\Delta(z) = \Omega_M(1+z)^3 + \Omega_\Lambda$, $D_H$ is the
Hubble distance, $\Omega_M$ is the dimensionless matter density and
$\Omega_{\Lambda}$ is the dimensionless cosmological constant density. 
All other quantities (masses, lengths and frequencies) in Eq.~\eqref{strain} should be instead be interpreted as measured by an observer in the source frame. 

Alternatively, one might wish to use quantities measured by an observer at the detector's location to compute the strain amplitude of Eq.~\eqref{strain}. Detector-frame quantities are related to source-frame ones by
powers of $(1+z)$, namely all quantities with dimensions
$[$mass$]^p$ (in our geometrized units $G=c=1$) are multiplied by the factor  $(1+z)^p$, e.g. masses are multiplied by $(1+z)$ (``redshifted masses''), frequencies are divided by the same factor (``redshifted frequencies''), while the comoving distance is multiplied by a factor $(1+z)$, thus becoming the luminosity distance $D_L=D_c(1+z)$. Since the  strain amplitude of Eq.~\eqref{strain} is dimensionless, 
that equation yields the same result when using detector-frame quantities as when using source-frame ones.

The typical distance up to which BH-condensate sources are detectable
can be estimated by defining an ``angle-averaged range''
$D_{\rm range}$ as the luminosity distance at which either the SNR
$\rho(D_{\rm range})=8$ [cf. Eq.~\eqref{SNR_Enrico}] for coherent
searches, or $h( D_{\rm range})=h_{\rm thr}$ for semicoherent searches
[cf. Eq.~\eqref{hthr}].

In Fig.~\ref{reachradius} we show
$D_{\rm range}$ for both LISA and LIGO at design sensitivity under
different assumptions on the initial BH spin.
The left panels refer to single coherent observation with
$T_{\rm obs}=4\,{\rm yr}$ for LISA ($T_{\rm obs}=2\,{\rm yr}$ for
Advanced LIGO), whereas the right panels refer to a (presumably more
realistic) semicoherent search with ${\cal N}=121$ coherent segments
of duration $T_{\rm coh}=250\,{\rm hr}$. In the more optimistic case,
sources are detectable up to cosmological distances of
$\sim 20\,{\rm Gpc}$ ($\sim 2\,{\rm Gpc})$ if the BH is nearly
extremal and the boson mass is in the optimal mass range
$m_s\sim 10^{-17}\,{\rm eV}$ ($m_s\sim 10^{-13}\,{\rm eV}$) for LISA
(LIGO). For the semicoherent search, $D_{\rm range}$ is reduced by
roughly one order of magnitude, with a maximum detector reach
$\sim 2\,{\rm Gpc}$ and $\sim 200\,{\rm Mpc}$ for LISA and Advanced
LIGO, respectively.

\section{Massive black hole population models}
\label{sec:SMBHs}

An assessment of the detectability of GWs from superradiant
instabilities requires astrophysical models for the massive BH population.
In this section we describe the models 
adopted in our study, and in particular our assumptions on {\bf (A)}
the mass and spin distribution of isolated massive BHs, {\bf (B)}
their Eddington ratio distribution, and {\bf (C)} their merger
history.

\subsection{\bf Mass and spin distribution of isolated black holes}\label{massdist}
Let $n$ be the comoving-volume number density of BHs. For the mass and
spin distribution of isolated BHs we consider:

\begin{itemize}
\item[{\bf (A.1)}] A model where ${d^2 n}/(d \log_{10} M d \chi)$ is
computed using the semianalytic galaxy formation model of
\cite{Barausse:2012fy} (with later improvements described in
\cite{Sesana:2014bea,newpaper,letter}). This distribution is
redshift-dependent and skewed toward large spins, at least at low
masses (cf.~\cite{Sesana:2014bea}). It also has a negative slope
${\rm d}{n}/{{\rm d}\log_{10} M}\propto M^{-0.3}$ for BH masses
$M<10^7 M_\odot$, which is compatible with observations
(cf.~\cite{Sesana:2014bea}, Figure 7). The normalization is calibrated
so as to reproduce the observed $M$--$\sigma$ and $M$--$M_{\star}$
scaling relations of \cite{Kormendy:2013dxa}, where $\sigma$ is the
galaxy velocity dispersion and $M_{\star}$ is the stellar mass. We
also account for the bias due to the resolvability of the BH sphere of
influence~\cite{Shankar:2016yrf,Barausse:2017uyr}. Because of the
slope, normalization and spin distribution, this model is
\textit{optimistic}.

\item[{\bf (A.2)}] An analytic mass function~\cite{Babak:2017tow,Gair:2017ynp} \begin{equation}
  \diff{n}{\,\log_{10} M} = 0.005 \left( \frac{M}{3\times 10^6 \msun} \right)^{-0.3}~\mathrm{Mpc^{-3}},
  \label{mfapprox}
\end{equation}
which we use for redshifts and BH masses in the range $10^4 M_\odot< M<10^7 M_\odot$ and $z<3$. For $M>10^7 M_\odot$ we use a mass distribution with normalization $10$ times lower than the optimistic one.  For this model we use a \textit{uniform} distribution of the initial spins $\chi\in [0,1]$. Because of the lower
normalization and the spin distribution, this model is \textit{less optimistic}. 

\item[{\bf (A.3)}] An analytic mass function \begin{equation}
\diff{n}{\,\log_{10} M} = 0.002 \left( \frac{M}{3\times 10^6 \msun} \right)^{0.3}~\mathrm{Mpc^{-3}},
  \label{mfpess}
\end{equation}
which we use again for $10^4 M_\odot< M<10^7 M_\odot$ and $z<3$, whereas for $M>10^7 M_\odot$ we use a mass distribution with normalization $100$ times lower than the optimistic one.  For this model we also consider a \textit{uniform} distribution of the initial spins $\chi\in [0,1]$. Because
of the normalization, slope and spin distribution, this model is
\textit{pessimistic}.

\end{itemize}

\subsection{Black hole mergers} \label{sec:mergers}

Our standard choice for BH mergers is to compute the comoving-volume
number density $n_m$ of mergers per (logarithmic) unit of total mass
$M_{\rm tot}= M_1+ M_2$, unit redshift and
(logarithmic) unit of mass ratio $q=M_2/M_1\leq1$, i.e.
\begin{equation}
\nu(M_{\rm tot},z,q)\equiv \frac{d^3 n_{\rm m}}{d \log_{10} M_{\rm
    tot}dz d \log_{10} q}\,,
\end{equation}
from the semianalytic model of \cite{Barausse:2012fy}.

We can then estimate the average number of mergers (between $z$ and
$z+dz$) for a BH of mass $M$
as 
\begin{equation}
dN_m(M,z)=\frac{\mu(M,z)}{\phi(M,z)} dz\,.
\end{equation} 
Here 
\begin{equation}
  \phi(M,z)\equiv
  \diff{n}{\,\log_{10} M} = \int \frac{ {d} ^2 n}{
    {d} \log_{10} M {d} \chi}  {d} \chi
\end{equation} 
is the (isolated BH) mass function, and
\begin{align}&\mu(M_{\rm tot},z)\equiv \frac{ {d}^2 n_{\rm
      merger}}{ {d} \log_{10} M_{\rm tot} {d}z}=\int_{q>q_c}
  \nu\, {d}\log_{10} q\,,\notag\\\notag
\end{align} 
where $q_c$ is the critical mass ratio above which we assume mergers
make an impact. In practice, most BH mergers in our
semianalytic models have $q\gtrsim0.01$--$0.001$ (especially in the LISA band, cf.~\cite{Dvorkin:2017vvm}), so our results
are robust against the exact choice of $q_c$. 
Nevertheless, to be on the conservative side, we set $q_c=0$. 
A larger $q_c$ would produce a slightly lower BH merger number and, in
turn, a slightly higher number of boson-condensate sources, under the
conservative assumption that mergers destroy the boson cloud.
We can then compute the average number of mergers experienced by a BH
of mass $M$ in the redshift interval $[z_1,z_2]$ as
\begin{equation}
N_m=\int_{z_1}^{z_2} \diff{N_m}{z} {d} z\,.
\end{equation}

Note that the number of mergers depends on the seeding mechanisms of
the massive BH population, as well as on the ``delays'' between the
mergers of galaxies and the mergers of the BHs they host
[cf. e.g. \cite{Klein:2015hvg}]. 

When computing the average number of mergers $N_m$ to be used to
estimate the number of boson-condensate GW events from \textit{isolated} BHs,
i.e. when evaluating the number of resolved events [Eq.~\eqref{RATES}
below] and the amplitude of the stochastic background
[Eq.~\eqref{stochastic1} below], we consider the ``popIII'' model of
\cite{Klein:2015hvg} (a light-seed scenario with delays). Choosing a
different seed model would not alter our conclusions.
However, when considering the constraints that can be placed on the
boson mass by direct observations of BH \textit{coalescences} by LISA, we
consider all three models presented in~\cite{Klein:2015hvg}
(``popIII'', ``Q3'' and ``Q3nod''). These models correspond
respectively to light seeds with delays between a galaxy merger and
the corresponding binary BH merger; heavy seeds with delays; and heavy
seeds with no delays; and they are chosen to bracket the theoretical
uncertainties on the astrophysics of BH seed formation and BH delays.

\subsection{Accretion}
Clearly, accretion is competitive with the superradiant extraction of angular momentum from the BH~\cite{Brito:2014wla}, so it is important to quantify its effect. We estimate the accretion time scale via the
Salpeter time, 
\begin{equation}
\label{salpeter}
t_S=4.5\times 10^8 {\rm\, yr\,} \frac{\eta}{f_{\rm Edd} (1-\eta)}\,,
\end{equation}
where $f_{\rm Edd}$ is the Eddington ratio for mass accretion, and the
thin-disk radiative efficiency $\eta$ is a function of the spin related to the specific
energy $E_{_{\rm ISCO}}$ at the innermost stable circular
orbit~\cite{Bardeen:1972fi}:
\begin{eqnarray}
&&\eta=1-E_{_{\rm ISCO}}\,,\\
&&{E}_{_{\rm ISCO}}=\sqrt{1-\frac{2}{3{r}_{_{\rm ISCO}}}}\,,\\ 
&&{r}_{_{\rm ISCO}}=3+Z_{2}-\frac{\chi}{|\chi|}\sqrt{(3-Z_{1})(3+Z_{1}+2Z_{2})}\,,\\ 
&&Z_{1}=1+(1-\chi^2)^{1/3}\left[(1+{\chi})^{1/3}+(1-{\chi})^{1/3}\right]\,,\\ 
&&Z_{2}=\sqrt{3{\chi}^{2}+Z_{1}^{2}}\,.
\end{eqnarray}

For the Eddington ratio $f_{\rm Edd}$ we consider three models:

\begin{itemize}

\item[{\bf (C.1)}] We use the results of our semianalytic model to
construct probability distribution functions for $f_{\rm Edd}$ at
different redshifts and BH masses.

\item[{\bf (C.2)}] We adopt a simple model in which $f_{\rm Edd}=1$
for $10$\% of the massive BHs, and $f_{\rm Edd}=0$ for the remaining
ones. (The choice of 10\% is a reasonable estimate for the duty cycle
of active galactic nuclei~\cite{Shankar:2011mc,2016ApJ...831..203P}).

\item[{\bf (C.3)}] Finally, we consider a \textit{very pessimistic
  model} in which all BHs have $f_{\rm Edd}=1$. Although unrealistic,
this models maximizes the effects of accretion, and therefore it
yields the most conservative lower bound for the superradiant
instability time scale.

\end{itemize}

\section{Stellar mass black hole population models}\label{sec:Irina}

We now turn to a description of stellar-mass BHs, which are of interest for
LIGO. Here we have to model {\bf (A)} extragalactic BHs, which turn out
to dominate the stochastic background of GWs from ultralight bosons,
and {\bf (B)} Galactic BHs, which (as pointed out
in~\cite{Arvanitaki:2014wva,Arvanitaki:2016qwi}) are dominant in terms
of resolvable signals.

\subsection{Extragalactic BHs}\label{sec:extragal}

In the standard scenario, stellar-mass BHs are the end products of the
evolution of massive ($M\gtrsim 20M_{\odot}$) stars. They form either
via direct collapse of the star or via a supernova explosion followed by
fallback of matter (failed supernova). This process depends on various
parameters, such as stellar metallicity, rotation and interactions
with a companion if the star belongs to a binary
system~\cite{2009A&A...497..243D,2014ApJ...789..120B,2016MNRAS.458.2634M,2016A&A...588A..50M}. In
particular, the metallicity of the star determines the strength of
stellar winds and can thus have a significant impact on the mass of
the stellar core prior to
collapse~\cite{2008NewAR..52..419V,2010ApJ...715L.138B}. In addition,
BHs can grow hierarchically through multiple mergers that occur in
dense stellar clusters
\cite{2016ApJ...831..187A,Fishbach:2017dwv,Gerosa:2017kvu,2017ApJ...834...68C}. This
process is expected to leave an imprint on the distribution in the
mass-spin plane: while BHs grow in mass via mergers their spins
converge to values around $\sim 0.7$ with little or no support below
$\sim 0.5$ \cite{Berti:2008af,Fishbach:2017dwv,Gerosa:2017kvu}.

In this work we consider only BH formation from core collapse of
massive stars. We use the analytic fits for the BH mass as a function
of initial stellar mass and metallicity
from~\cite{2012ApJ...749...91F}, embedded in the semianalytic galaxy
evolution model from~\cite{2016MNRAS.461.3877D}. In particular, the
latter model describes the production of metals by
stars~\cite{1995ApJS..101..181W} and the evolution of the metallicity
of the interstellar medium, which is inherited by the stars that form
there. The extragalactic BH formation rate as a function of mass and
redshift reads
\begin{equation}
\frac{d \dot{n}_{\rm eg}}{dM}=\int \textrm{d}{\cal M_\star}\psi[t-\tau({\cal {M_\star}})]\phi({\cal M_\star})\delta[{\cal M_\star}-g^{-1}({M})]\,,
\label{Irina}
\end{equation}
where $\tau({\cal M_\star})$ is the lifetime of a star with mass
${\cal M_\star}$, $\phi({\cal M_\star})$ is the stellar initial mass
function, $\psi(t)$ denotes the cosmic star formation rate (SFR) density
and $\delta$ is the Dirac delta. We use the fit to the
cosmic SFR described in~\cite{2015MNRAS.447.2575V}, calibrated 
to observations~\cite{2011ApJ...737...90B,2013ApJ...770...57B}. We
adopt a Salpeter initial mass function
$\phi({\cal M_\star})\propto {\cal
  M_\star}^{-2.35}$~\cite{1955ApJ...121..161S} in the mass range
${\cal M_\star}\in[0.1-100]\,M_{\odot}$ and use the stellar lifetimes
from~\cite{2002A&A...382...28S}. The initial stellar mass
$\cal M_\star$ and BH mass $M$ are related by the function
$M=g({\cal M_\star})$, which can be (implicitly) redshift-dependent
(through its dependence on stellar metallicity), and which
we take from the ``delayed'' model of~\cite{2012ApJ...749...91F}.

\subsection{Galactic BHs}\label{sec:gal}

Resolvable signals are expected to be dominated by Galactic
stellar-mass BHs~\cite{Arvanitaki:2014wva}. We estimate the 
present-day mass function of these BHs as
\begin{equation}\label{MWBHs}
\frac{dN_{\rm MW}}{d M}=\int dt\frac{{\rm SFR}(z)}{{\cal M_\star}}\frac{dp}{d{\cal M_{\star}}}\left\vert\frac{dM}{d{\cal M_\star}}\right\vert^{-1}\,, 
\end{equation}
where $N_{\rm MW}$ denotes the number of BHs in the Galaxy, ${dp}/{d{\cal M_{\star}}}$ is the normalized Salpeter initial mass
function (i.e. the probability of forming a star with mass between
${\cal M_\star}$ and ${\cal M_\star}+d{\cal M_\star}$), and ${\rm SFR}(z)$ denotes the SFR of
Milky-Way type galaxies as a function of $z$~\cite{2013ApJ...770...57B,2013ApJ...762L..31B}. The integration is over all cosmic times till the present epoch.
The (differential) relation between BH mass and
initial stellar mass $dM/{d{\cal M_\star}}$ is taken from the ``delayed'' model
of~\cite{2012ApJ...749...91F}, 
and is also a function of redshift via the metallicity. For the latter,
we use the results of~\cite{2016MNRAS.456.2140M} to describe its evolution with cosmic time.
We then ``spread'' ${dN_{\rm MW}}/{d M}$ throughout the Galaxy in order to obtain a (differential)
density ${dn_{\rm MW}}/{d M}$, by assuming that
the latter is everywhere proportional to the (present)
stellar density. To this purpose, we describe the Galaxy by a bulge+disk
model, where the bulge follows a Hernquist
profile~\cite{1990ApJ...356..359H} with mass
$\sim 2\times 10^{10} M_\odot$ and scale radius
$\sim 1\,{\rm kpc}$~\cite{Shen:2003sda}, and the disk is described by
an exponential profile with mass $\sim 6\times 10^{10} M_\odot$ and
scale radius $\sim 2 $ kpc~\cite{1998A&A...330..136P}.

Since these models (for both Galactic and extragalactic BHs) do not predict the initial BH spins, we
assume a uniform distribution and explore different ranges (from optimistic to pessimistic):
$\chi\in [0.8,1]$, $[0.5,1]$, $[0,1]$ and $[0,0.5]$.

\section{Event rates for LISA and LIGO}\label{sec:rates}

Having in hand the calculation of the GW signal of
Sec.~\ref{sec:bosonGW} and the astrophysical models of
Secs.~\ref{sec:SMBHs} and~\ref{sec:Irina}, we can now compute event
rates for LISA and LIGO. We consider two separate classes of sources:
{\bf (A)} boson-condensate GW events which are loud enough to be
individually resolvable, and {\bf (B)} the stochastic background of
unresolvable sources.

\subsection{Resolvable sources}\label{rates:detectable}

In the limit in which the (detector-frame) signal duration
$\Delta t (1+z)$ is small compared to the observation time
$T_{\rm obs}$, $\Delta t (1+z) \ll T_{\rm obs}$, the number of
resolvable events is proportional to the observation
time~\cite{Hartwig:2016nde}:
\begin{equation}\label{RATES1}
{N}=  T_{\rm obs} \int_{\rho >8}\frac{d^2 \dot{n}}{d M  d \chi} \frac{dt}{dz} 4 \pi D_c^2 dz d M d \chi \,,
\end{equation}
where 
\begin{equation}
\frac{{\rm} d t}{{\rm} d z}=\frac{1}{H_0 \sqrt{\Delta} (1+z)}
\end{equation} 
is the derivative of the lookback time with respect to redshift.

For long-lived sources with detector-frame duration
$\Delta t (1+z) \gg T_{\rm obs}$, the number of detections does
\textit{not} scale with the observation time, but rather with the
``duty cyle'' $\Delta t/t_f$, where $t_f\equiv n/\dot{n}$ is the
formation time scale of the boson condensate. For example, if BHs form
a boson condensate only once in their cosmic history, $t_f$ is the age
of the Universe $t_0\approx 13.8\,{\rm Gyr}$. This duty cycle has the
same meaning as the duty cycle of active galactic nuclei: it accounts
for the fact that, at any given time, only a fraction of the BH
population will be emitting GWs via boson condensates. Because of the
ergodic theorem, this fraction is given by the average time fraction
during which a BH emits GWs via boson condensates. This average time
fraction is indeed the duty cycle $\Delta t/t_f$. Therefore, the
number of resolved sources when $\Delta t (1+z) \gg T_{\rm obs}$ is
simply
\begin{align}
{N}&=  \int_{\rho>8}\frac{d^2 {n}}{d M  d \chi} \frac{\Delta t}{t_f} \frac{d V_c}{dz}dz d M d \chi \nonumber\\
&= \int_{\rho >8}\frac{d^2 \dot{n}}{d M  d \chi} {\Delta t} \frac{d V_c}{dz}dz d M d \chi \,,\label{RATES2}
\end{align}
where $dV_c=4 \pi D_c^2 {d D_c}$.

Equations~\eqref{RATES1} and \eqref{RATES2} can be merged into a
single expression that remains valid also in the intermediate regime
$\Delta t(1+z) \sim T_{\rm obs}$. Indeed, the probability that a
signal lasting a time span $\Delta t (1+z)$ (in the detector frame)
overlaps with an observation of duration $T_{\rm obs}$ is simply
proportional to the sum of the two durations,
$\Delta t(1+z) + T_{\rm obs}$. This can be understood in simple
geometric terms: for the signal to overlap with the observation window
(which we define, without loss of generality, to extend from $t=0$ to
$t=T_{\rm obs}$), the signal's starting time should fall between
$t=-\Delta t(1+z)$ and $t=T_{\rm obs}$, i.e. in a time interval of
length $\Delta t(1+z) + T_{\rm obs}$. Therefore, we can estimate the
number of observable GW events as
\begin{equation}\label{RATES}
{N}=  \int_{\rho >8}\frac{d^2 \dot{n}}{d M  d \chi} \left(\frac{T_{\rm obs}}{1+z}+\Delta t\right) \frac{d V_c}{dz}dz d M d \chi \,.
\end{equation}
Since $d D_c/dz=(1+z)dt/dz$, it can be easily checked this equation
reduces to Eqs.~\eqref{RATES1} and \eqref{RATES2} in the limits
$\Delta t (1+z) \ll T_{\rm obs}$ and $\Delta t (1+z) \gg T_{\rm obs}$,
respectively.

For extragalactic LIGO sources we compute ${d^2 \dot{n}}/{d M d \chi}$ from the
astrophysical models of Sec.~\ref{sec:extragal}, while for LISA and galactic LIGO sources we
compute ${d^2 {n}}/{d M d \chi}$ as described in Secs.~\ref{sec:SMBHs} and~\ref{sec:gal}
and then assume
${d^2 \dot{n}}/{d M d \chi}=({d^2 {n}}/{d M d \chi})/t_0$. This
corresponds to assuming that the boson-condensate formation time
$t_f=t_0$ equals the age of the Universe, or that BHs radiate via
boson condensates only once in their lifetime. This conservative
assumption does not affect our results very significantly. Once a
BH-boson system radiates, its spin decreases to low values, while the
mass remains almost unchanged. For the BH to emit again via boson
condensates, its spin must grow again under the effect of accretion or
mergers. In this process, however, the BH mass also grows rapidly: for
example, the simple classic estimates by Bardeen~\cite{Bardeen:1970zz}
imply that when a BH spins up from $\chi=0$ to $\chi=1$ via accretion,
its mass increases by a factor $\sqrt{6}$. So even if new boson clouds
form due to the instability of higher-$m$ modes, the instability time scales will be much larger [cf. Eq.~\eqref{omega}] and the GW flux will be highly suppressed [cf. Ref.~\cite{Yoshino:2013ofa}].

\begin{figure}[t]
\begin{center}
\epsfig{file=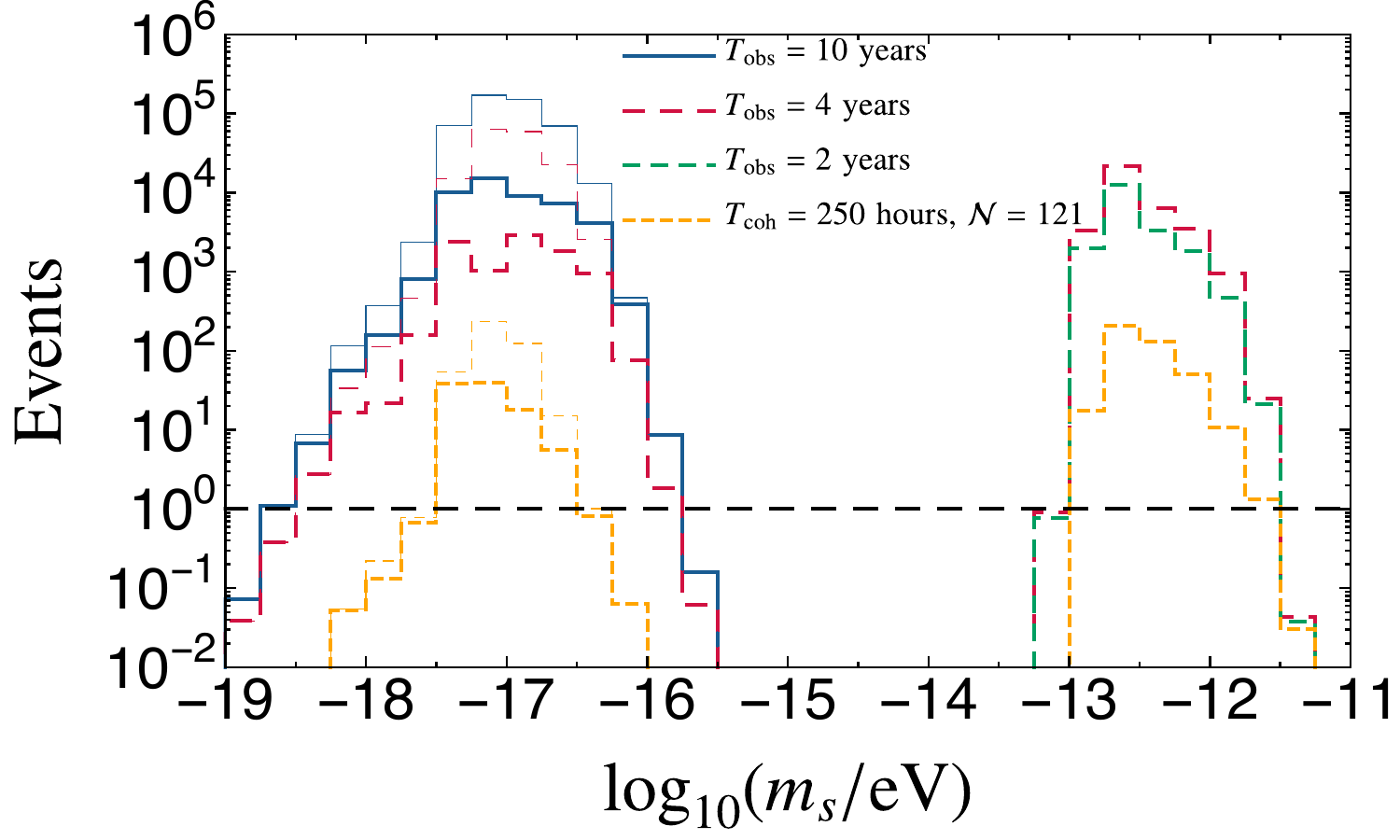,width=\columnwidth,angle=0,clip=true}
\caption{Number of resolved LIGO and LISA events for our optimistic BH
  population models as a function of the boson mass with different
  observation times $T_{\rm obs}$, using both full and semicoherent
  searches. Thick (thin) lines were computed with (without) the
  confusion noise from the stochastic background.
\label{events_all}}
\end{center}
\end{figure}

\begin{figure*}[htb]
\begin{center}
\begin{tabular}{c}
\epsfig{file=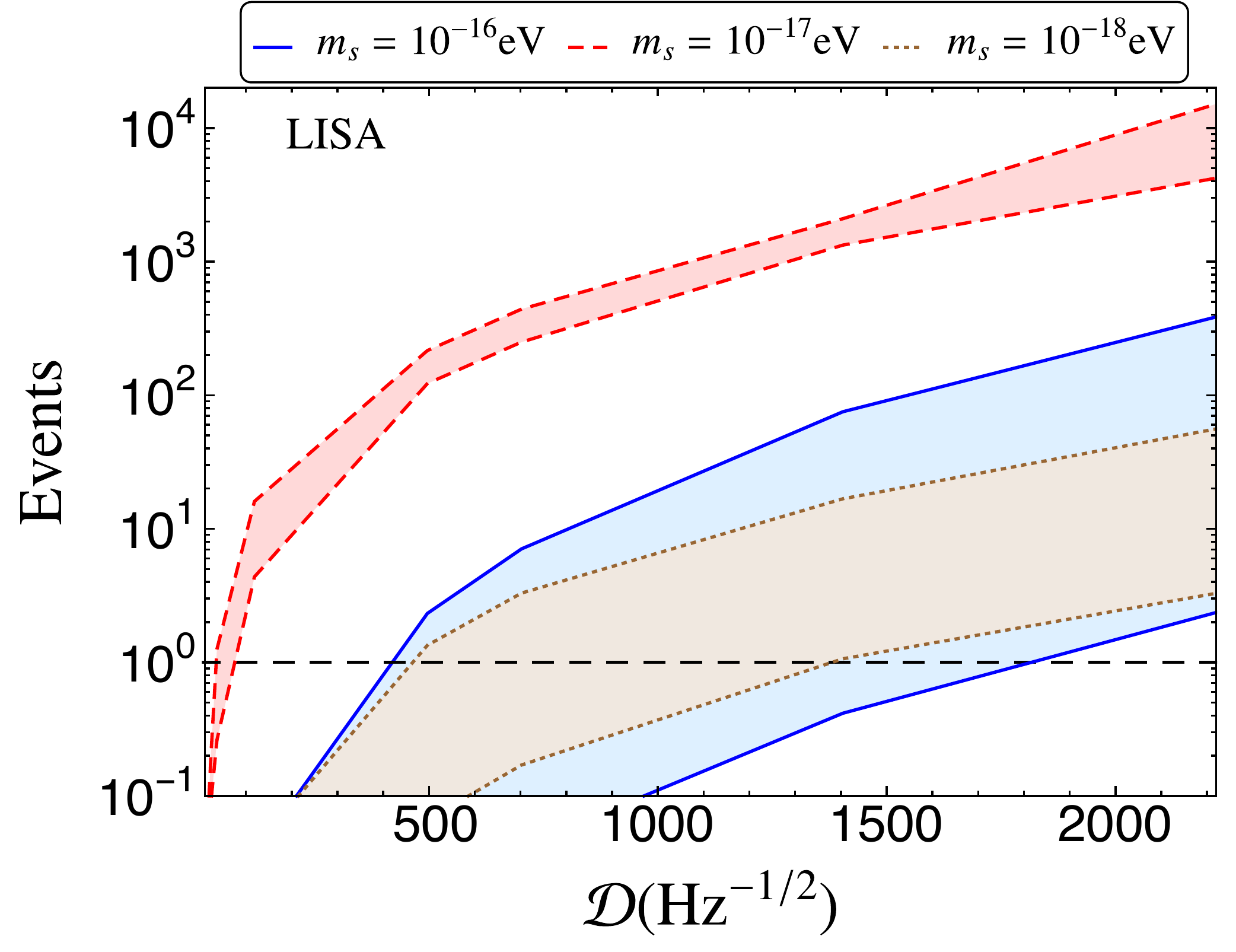,width=\columnwidth,angle=0,clip=true}
\epsfig{file=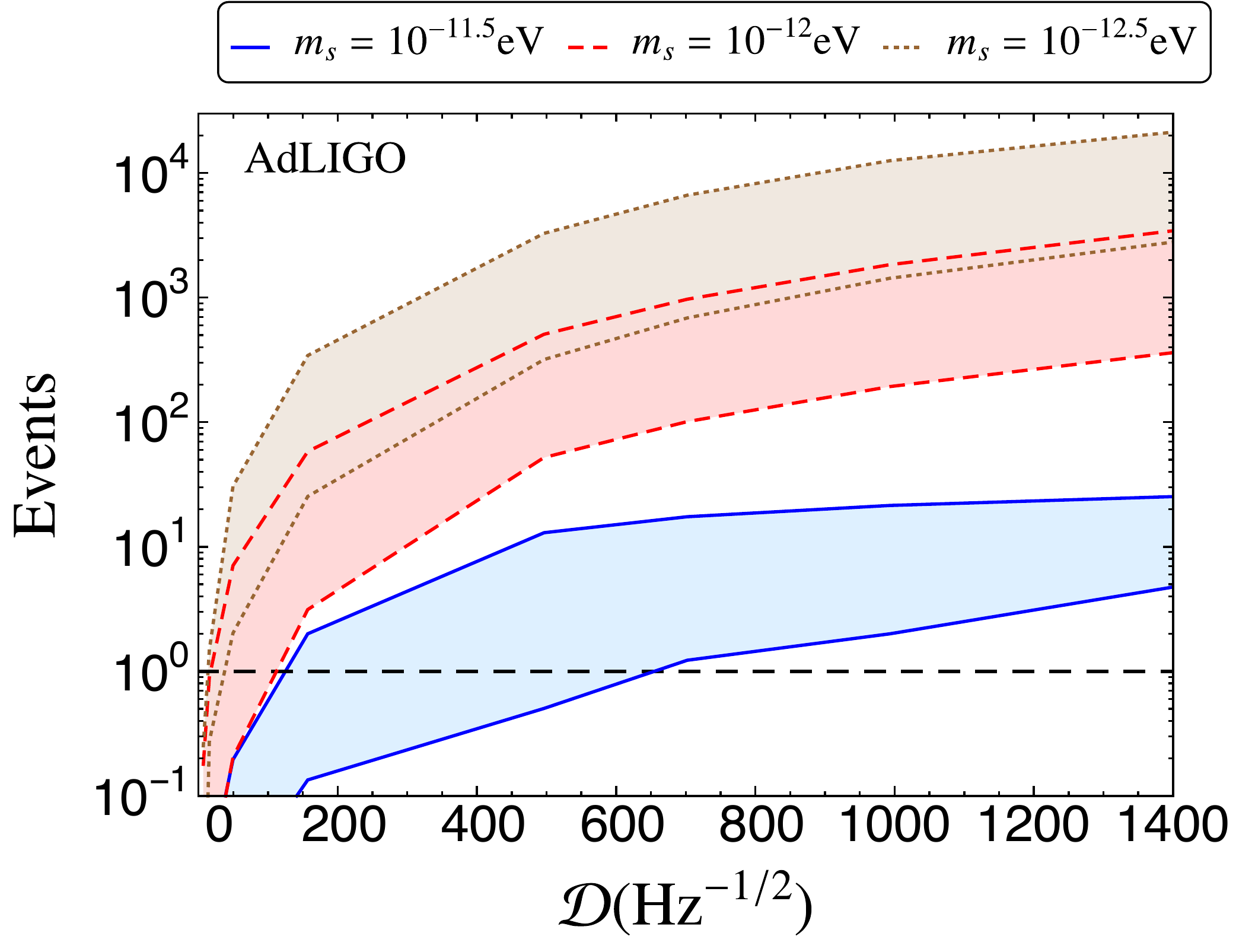,width=\columnwidth,angle=0,clip=true}
\end{tabular}
\caption{Left: Number of events as a function of the sensitivity depth
  $\mathcal{D}$ [Eq.~\eqref{SenDep}] for selected boson masses in the
  LISA band and accretion model $\textit{\bf (C.1)}$. The bottom (top)
  of each shadowed region correspond to the pessimistic (optimistic)
  model. Right: Same, but for boson masses in the LIGO band. Here the
  bottom (top) of each shadowed region correspond to pessimistic
  (optimistic) spin distributions.\label{events_depth}}
\end{center}
\end{figure*}

Our main results for resolvable rates are summarized in
Fig.~\ref{events_all}, Fig.~\ref{events_depth},
Table~\ref{tab:events_LISA} and Table~\ref{tab:events_LIGO}.

In Fig.~\ref{events_all} we focus on optimistic models and we show
how the number of individually resolvable events depends on the
observation time and on the chosen data-analysis method. More
specifically, for LISA we use the BH mass-spin distribution
model {\bf (A.1)} and accretion model {\bf (C.1)}, while for LIGO we
adopt the optimistic spin distribution $\chi_i\in [0.8,1]$.
We bracket uncertainties around the nominal LISA mission duration of
$T_{\rm obs}=4 \,{\rm yr}$~\cite{Audley:2017drz} by considering single
observations with duration $T_{\rm obs}=(2,\,4,\,10)\,{\rm yr}$. We
also show rates for a (presumably more realistic) semicoherent search
with 121 segments of $T_{\rm coh}=250\,{\rm hours}$ coherent
integration time\footnote{The number of resolved events for other choices of number of segments and coherent integration time can be obtained from Fig.~\ref{events_depth} and expressing the sensitivity depth as $\mathcal{D}\approx T_{\rm coh}^{1/2}{\cal N}^{1/4} 25^{-1}$ [cf. Eqs.~\eqref{hthr} and~\eqref{SenDep}].}. For Advanced LIGO at design sensitivity, we similarly
consider single observations lasting either $T_{\rm obs}=2 \,{\rm yr}$
or $T_{\rm obs}=4 \,{\rm yr}$, as well as a semicoherent search with
121 segments of $T_{\rm coh}=250\,{\rm hours}$ coherent integration
time.

Figure~\ref{events_all} (together with Figure~3 in~\cite{Brito:2017wnc}) shows
that the number of resolvable events is strongly dependent on the
boson mass and on the astrophysical model.  

For LISA, our astrophysical populations contain mostly BHs in the mass
range $10^4 M_\odot < M < 10^8 M_\odot$, and the sensitivity curve
peaks around a frequency corresponding to $m_s\sim 10^{-17} {\rm eV}$
[cf. Fig.~1 of~\cite{Brito:2017wnc}]. These considerations -- together with the
condition for having an efficient superradiant instability (namely,
$M\mu\sim 0.4$ at large spin) -- translate into the range
$3\times 10^{-18}\,{\rm eV} \lesssim m_s \lesssim 5\times
10^{-17}\,{\rm eV}$ for the mass of detectable bosonic particles in a
semicoherent search.

For LIGO, our models predict that most BHs will be in the mass range
$3 M_\odot < M < 50 M_\odot$, and the most sensitive frequency band
corresponds to $m_s\sim 3\times 10^{-13} {\rm eV}$ [cf. Fig.~1
of~\cite{Brito:2017wnc}], translating into the range
$2\times 10^{-13}\,{\rm eV} \lesssim m_s \lesssim 3\times
10^{-12}\,{\rm eV}$ for the mass of detectable bosonic particles. 

In order to quantify the ``self-confusion'' noise due to the
stochastic background produced by BH-boson systems, in
Fig.~\ref{events_all} we also display the number of resolved events
that we would obtain if we omitted the confusion noise from the
stochastic background (cf. Fig.~1 of~\cite{Brito:2017wnc} and
Sec.~\ref{rates:stochastic}). Neglecting the confusion noise would
overestimate the number of resolvable events in LISA by one or two
orders of magnitude.

The rates computed in Figure~\ref{events_all} refer to our optimistic
astrophysical models. As shown in~\cite{Brito:2017wnc}, resolvable event rates
in the most pessimistic models are about one order of magnitude lower.
Nevertheless, it is remarkable that {\em even in the most pessimistic
  scenario} for direct detection (i.e., unfavorable BH mass-spin
distributions and semicoherent search method for the signal), bosonic
particles with $m_s\sim 10^{-17}\,{\rm eV}$
($m_s\sim 10^{-12}\,{\rm eV}$) would still produce around $5$ ($15$)
direct LISA (LIGO) detections of boson-condensate GW events.

In Figure~\ref{events_depth} we show how the number of events grows
with the sensitivity depth of the search ~\cite{Behnke:2014tma}, as
defined in Eq.~(\ref{SenDep}).  For LISA the number of events grows
roughly with $\mathcal{D}^3$, corresponding to $T_{\rm
  obs}^{3/2}$. This is expected from the fact that the number of
events for sources at $\gtrsim 30$ Mpc should grow with the sensitive
volume, and thus decrease with $\rho_{\rm crit}^{-3}$, where $\rho_{\rm crit}$ is the critical SNR for detection~\cite{Abadie:2010cf}. 

On the other hand, LIGO will be mostly sensitive to signals within the Galaxy. 
For a given boson mass and distance, $\tau_{\rm GW}\sim h^{-2}$ and $M\sim h^{1/8}$ [cf. Eqs.~\eqref{signalduration} and~\eqref{waveform}]. Since the Galactic stellar BH population obtained from Eq.~\eqref{MWBHs} is well fitted by $dN/M\sim e^{-0.2 M}$, for a fixed volume the integral in Eq.~\eqref{RATES2} goes as 
\be
N\sim \int_{h >h_{\rm thr}} h^{-23/8} e^{-0.2 h^{1/8}} dh\sim h_{\rm thr}^{-15/8}\,,
\ee
where in the last step we took the leading order of the integral for small $h_{\rm thr}$. From Eq.~\eqref{SNR_Enrico} one has $h_{\rm thr}\propto T_{\rm obs}^{-1/2}$ and therefore $N\propto T_{\rm obs}^{15/16}$. This is in agreement with the scaling that we find.

Assuming the sensitivity depth of the last {\sc Einstein@Home} search
$\mathcal{D}\approx 35\rm{Hz}^{-1/2}$~\cite{TheLIGOScientific:2016uns}
and an optimal boson mass around $m_s\sim 10^{-12.5}$~eV, we find that
O1 should have detected 5 resolvable events for the optimistic spin
distribution $\chi\in [0.8,1]$, and 2 events for a uniform spin
distribution $\chi\in [0,1]$. As pointed out in~\cite{Brito:2017wnc}, these
optimal boson masses may already be ruled out by upper limits from
existing stochastic background searches~\cite{Brito:2017wnc}.  On the other
hand, the pessimistic spin distribution $\chi\in [0,0.5]$ is still
consistent with (the lack of) observations of resolvable BH-boson
GW events in O1, though marginally ruled out by the O1 stochastic background upper limits~\cite{Brito:2017wnc}.

Our results for resolvable event rates using different search
techniques, mass/spin and accretion models are summarized in
Tables~\ref{tab:events_LISA} and~\ref{tab:events_LIGO}. For LISA we
included ``self-confusion'' noise in our rate estimates, and using
different accretion models does not significantly affect our
results. Interestingly, even though the accretion models \textit{\bf
  (C.2)} and \textit{\bf (C.3)} are more pessimistic than model
\textit{\bf (C.1)}, they predict a slightly larger number of
resolvable events for boson masses in the optimal range around
$10^{-17}$ eV. This is because the self-confusion noise is lower for
models \textit{\bf (C.2)} and \textit{\bf (C.3)}
[cf. Section~\ref{rates:stochastic}], and thus the loss in signal is
more than compensated by the lower total (instrumental and
self-confusion) noise floor.

\begin{table}[th]
\begin{center}
\begin{tabular}{cc|cc}
$m_s [{\rm eV}]$ & Search method & Accretion model & Events \\
\hline 
\hline
$10^{-16}$  & Coherent & \textit{\bf (C.1)}  & 75~--~0\\
& Semicoherent & & 0\\
  &  Coherent &   \textit{\bf (C.2)}   & 75~--~0\\
& Semicoherent && 0\\
  &   Coherent & \textit{\bf (C.3)}   & 75~--~0\\
& Semicoherent & & 0\\
\hline
$10^{-17}$ & Coherent  & \textit{\bf (C.1)} & 1329~--~1022\\
& Semicoherent & & 39~--~5\\
     & Coherent &  \textit{\bf (C.2)} & 3865~--~1277\\
&  Semicoherent & & 36~--~4\\
    & Coherent &  \textit{\bf (C.3)} &  5629~--~1429\\
 & Semicoherent & & 39~--~5\\
\hline
$10^{-18}$  & Coherent  &  \textit{\bf (C.1)}  & 17~--~1  \\
& Semicoherent && 0\\
& Coherent  &     \textit{\bf (C.2)} &   18~--~1\\
& Semicoherent && 0\\
 & Coherent  &      \textit{\bf (C.3)} &  20~--~0\\
&  Semicoherent && 0\\
\end{tabular}
\end{center}
\caption{Number of resolvable events in the LISA band computed
  including the ``self-confusion'' noise from the stochastic
  background of BH-boson condensates for different accretion
  models. The lower and upper bounds correspond
  to the pessimistic and optimistic massive BH population models,
  respectively. For the semicoherent search we  use 121 segments of $T_{\rm
    coh}=250$ hours coherent integration time. For the coherent
  search, we adopt the nominal mission duration of $T_{\rm obs}=4$
  years.}
\label{tab:events_LISA}
\end{table}

\begin{table}[th]
\begin{center}
\begin{tabular}{ccc}
$m_s [{\rm eV}]$ & Search method  &  Events \\
\hline 
\hline
$10^{-11.5}$ & Coherent & 21~--~2\\
& Semicoherent & 1~--~0\\
\hline
$10^{-12}$  & Coherent & 1837~--~193\\
& Semicoherent & 50~--~2\\
\hline
$10^{-12.5}$ & Coherent  & 12556~--~1429\\
& Semicoherent & 205~--~15\\
\end{tabular}
\end{center}
\caption{Number of resolvable events for Advanced LIGO at design
  sensitivity. For the semicoherent search we use 121 segments of
  $T_{\rm coh}=250$ hours coherent integration time. For the coherent
  search, we set $T_{\rm obs}=2$ years. The lower and upper bounds
  correpond to the pessimistic  ($\chi \in [0,0.5]$) and optimistic
  ($\chi \in [0.8,1]$) spin distributions, respectively.}
\label{tab:events_LIGO}
\end{table}

\subsection{Stochastic background} \label{rates:stochastic}
 
In addition to individually resolvable sources, a population of
massive BH-boson condensates at cosmological distances can build
up a detectable stochastic background. This possibility is potentially
very interesting, given the spread in BH masses (and, hence, in boson
masses that would yield an instability) characterizing the BH
population at different redshifts, but to the best of our knowledge it
has not been explored in the existing literature.

The stochastic background can be computed from the formation rate
density per comoving volume $\dot{n}$ as~\cite{Phinney:2001di}
\begin{equation}
\label{stochastic1}
\Omega_{\rm gw}(f)=\frac{f}{\rho_c}\int_{\rho<8}  d \chi
d M d z \frac{d t}{d z}
\frac{d^2 \dot{n}}{d M d \chi}  \frac{d {E_s}}{d f_s}\,,
\end{equation}
where $\rho_c=3H_0^2/(8\pi G)$ is the critical density of the
Universe,
${d E_s}/{d f_s} $ is the energy spectrum in the source
frame, and $f$ is
the detector-frame frequency. Note that the integral is only over unresolved
sources with $\rho<8$.

For extragalactic stellar mass BHs (which are sources for LIGO), we
calculate $d^2 \dot{n}/d M d \chi$ based on the model of
Sec.~\ref{sec:Irina}, while for LISA sources we use the model of
Sec.~\ref{sec:SMBHs} to obtain $d^2 {n}/d M d \chi$, and then (as we
did for the resolved sources) we assume
$d^2 \dot{n}/d M d \chi= (1/t_0)( d^2 {n}/ d M d \chi)$. As before,
this corresponds to the conservative assumption that 
formation of boson condensates
 occurs only once in the cosmic history of each massive
BH.

We compute the energy spectrum as
\be
\frac{d E_s}{d f_s} \approx E_{\rm GW} \delta (f(1+z)-f_s)\,,
\ee
where we recall that $f_s$ is the frequency of the signal in the
source frame, $E_{\rm GW}$ is the total energy radiated by the boson
cloud in GWs during the signal duration $\Delta t$, and the Dirac
delta is ``spread out'' over a frequency window of size
$\sim \max[1/(\Delta t (1+z)),1/T_{\rm obs}]$ to account for the
finite signal duration and the finite frequency resolution of the
detector.  As in the calculation of the rates of resolved sources,
$\Delta t=\min\left(\tau_{\rm GW}, t_0\right)$ [cf. Eq.~\eqref{signalduration}] for LIGO sources, while we account for
mergers and accretion through Eq.~\eqref{deltat} for LISA sources.
Moreover, since our calculations rely on the implicit assumption that
the instability reaches saturation before GWs are emitted, our
estimates of the stochastic background only include BHs for which the
expected number of coalescences during the instability time scale is
$N_m<1$, and for which $\tau_{\rm inst}<\Delta t$ (which ensures that
the instability time scale is shorter than the merger and accretion
time scales).

The total energy emitted by the boson cloud during the signal duration
$\Delta t$ can be estimated by integrating the GW energy flux given by
Eq.~\eqref{gwflux}. Using Eq.~\eqref{Ms} we have
\be
\frac{d E_{\rm GW}}{dt}=\frac{d\tilde{E}}{dt}\frac{M_S^2}{M_f^2}=\frac{M_S^{\rm max}\tau_{\rm GW}}{\left(t+\tau_{\rm GW}\right)^2}\,,
\ee
and by integrating over a time $\Delta t$ we get
\be
E_{\rm GW}=\int_0^{\Delta t} dt \frac{d E_{\rm GW}}{dt}=\frac{M_S^{\rm max}\Delta t}{\Delta t+\tau_{\rm GW}}\,.
\ee
\begin{figure}[t]
\begin{center}
\begin{tabular}{c}
\epsfig{file=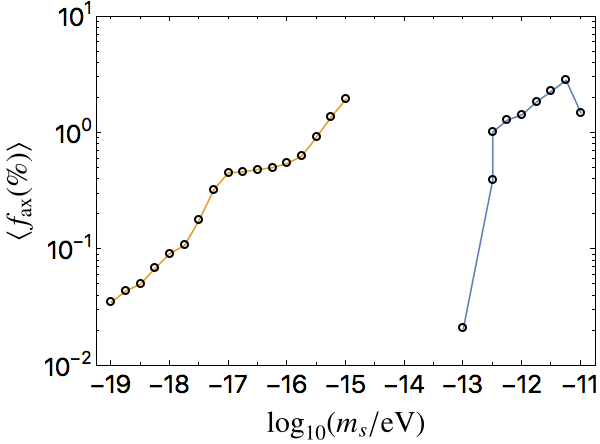,width=8cm,angle=0,clip=true}
\end{tabular}
\caption{Average fraction of mass of an isolated BH emitted by the bosonic cloud for the optimistic models.\label{faxion}}
\end{center}
\end{figure}

As shown in~\cite{Brito:2017wnc}, the order of magnitude of the stochastic
background can be estimated by computing the mass fraction of an
isolated BH that is emitted by the boson cloud through GWs. This can
be defined as
\be
f_{\rm ax}=\frac{E_{\rm GW}}{M_i}\,,
\ee
where we recall that $M_i$ is the initial mass of the BH. In
Fig.~\ref{faxion} we show the average $f_{\rm ax}$, weighted by the BH
population, for our most optimistic models. In the LIGO and LISA band
$f_{\rm ax}$ can be order ${\cal O}(1\%)$, leading to a very large
stochastic background~\cite{Brito:2017wnc}.

Note that Eq.~\eqref{stochastic1} cannot be applied to Galactic BHs
which emit in the LIGO band, because it implicitly assumes that the
number density of sources, $d^2 \dot{n}/ d M d \chi$, is homogeneous
and isotropic. That assumption is clearly invalid for Galactic BHs
[cf. Eq.~\eqref{MWBHs}]. However, in this case we can simply sum the
GW densities produced by the Galactic BH population at the position of
the detector. These densities are simply given by
$\rho_{\rm gw}=\dot{E}/(4 \pi r^2)=(5/4) \pi f_s^2 h^2$
[cf. Eq.~\eqref{waveform}], $r$ being the distance from the source to
the detector. (Note that we neglect redshift and cosmological effects,
since those are negligible inside the Galaxy.) Therefore, the GW
energy density per (logarithmic) unit of frequency coming from each BH
in the Galaxy is simply
$d\rho_{\rm gw}/d\ln f \approx (5/4) \pi f_s^2 h^2 \delta(\ln f-\ln
f_s)$, where the Dirac delta is ``spread out'' over a frequency window
of size $\sim \max[1/\Delta t,1/T_{\rm obs}]$ to account for the
finite duration of the signal and the finite frequency resolution of
the detector. Therefore, the contribution to the stochastic background
from the population of Galactic BHs can be written as
\begin{align}\label{stochastic2}
\Omega_{\rm gw}(f)&=\frac{1}{\rho_c}\int d M   d V\frac{d{n}_{\rm MW}}{d M}\frac{\Delta t}{t_0}\frac{d\rho_{\rm gw}}{d\ln f}\,.
\end{align}
Here $dV$ denotes a volume integration over the Galaxy, and
$\Delta t/{t_0}$ is again a duty cycle (i.e., we assume that Galactic
BHs emit via boson condensates only once in their cosmic history).

To compute the ${\rm SNR}$ for the stochastic background we use
\be\label{SNRstoch}
\rho_{\rm stoch}=\sqrt{T_{\rm obs}\int_{f_{\rm min}}^{f_{\rm max}}df \frac{\Omega^2_{\rm GW}}{\Omega^2_{\rm sens}}}\,.
\ee
For LISA we have~\cite{Cornish:2001qi}
\be
\Omega_{\rm sens}=S_h(f) \frac{2\pi^2}{3H_0^2}f^3\,,
\ee
while for LIGO~\cite{Allen:1997ad}
\be
\Omega_{\rm sens}=\frac{S_h(f)}{\sqrt{2}\Gamma_{IJ}(f)} \frac{2\pi^2}{3H_0^2}f^3\,,
\ee
where LIGO's noise power spectral density $S_h(f)$ is assumed to be
the same for both Livingston and Hanford, and $\Gamma_{IJ}$ is the
overlap reduction function as defined in~\cite{Thrane:2013oya}. Notice
the $1/\sqrt{2}$ factor in $\Omega_{\rm sens}$ for LIGO compared to LISA, due
to the use of data from two detectors instead of one.

As shown in Fig.~2 of~\cite{Brito:2017wnc},
the SNR for this stochastic signal can be very high. Since the galactic background only contributes to the full spectrum in a very narrow frequency window around $f_s$, the contribution of the extragalactic background to the SNR largely dominates. When computing
the background for LISA we assumed the semianalytic accretion model
$\textit{\bf (C.1)}$. Considering the most pessimistic accretion model
$\textit{\bf (C.3)}$ lowers the maximum SNR by at most a factor two.

\section{Excluding or measuring boson masses through LISA
  black hole spin measurements}
\label{sec:holes}

So far we have focused on the direct detection of GWs from bosonic
condensates. However it is also possible to infer the existence of
light bosons in an indirect way. As shown in Fig.~\ref{fig:Regge}, the
existence of a light boson would lead to the absence of BHs with spin
above the corresponding superradiant instability window (i.e., there
would be holes in the BH mass-spin ``Regge
plane''~\cite{Arvanitaki:2010sy}). In this section we show that LISA
measurements of the spins of merging massive BHs can be used to either
rule out bosonic fields in the mass range
$[4.5 \times 10^{-19}, 7.1\times 10^{-13}]$~eV, or even more
excitingly (if fields in the mass range $[10^{-17}, 10^{-13}]$~eV
exist in nature) to measure their mass with percent accuracy.

In principle we could carry out a similar analysis using astrophysical
models for stellar-mass BH binary mergers detectable by Advanced LIGO
or third-generation Earth-based detectors. However, spin magnitude
measurements for the components of a merging BH binaries are expected
to be poor ($\Delta \chi\sim 0.3$ at best) even with third-generation
detectors~\cite{Vitale:2016avz,Vitale:2016icu}. In addition, the mass range of BHs
detectable by LIGO or future Earth-based interferometers overlaps in
mass with existing spin estimates from low-mass X-ray binaries
(see~\cite{Remillard:2006fc,Reynolds:2013qqa,Miller:2014aaa,Middleton:2015osa}
for reviews of current BH spin estimates). In summary, we focus on
LISA for two main reasons: 

\begin{itemize}
\item[(i)] LISA allows for percent-level determinations of massive BH
  spins (see e.g. Fig.~9 of~\cite{Klein:2015hvg}).
\item[(ii)] In comparison with current electromagnetic estimates of
  massive BH spins, which can be used to exclude boson masses in
  the range $[10^{-20}, 10^{-17}]$~eV (see
  e.g.~\cite{Pani:2012vp,Baryakhtar:2017ngi}), LISA BH spin
  measurements can probe lower BH masses; therefore, depending on the
  details of massive BH formation models, they can exclude (or
  measure) boson masses all the way up to
  $m_s\sim 7\times 10^{-13}$~eV.
\end{itemize}

One of our main tasks in this context is to determine whether LISA
observations can distinguish between two models: one where a
massive boson exists (depleting the corresponding instability region
in the BH Regge plane) and a ``standard'' model where no depletion
occurs.  This is a standard Bayesian model selection problem (see
e.g.~\cite{Gair:2010bx,Sesana:2010wy,Berti:2011jz} for previous
applications of model selection to LISA observations of massive BH
binaries).

We simulate massive BH binary catalogs corresponding to the three
astrophysical models described in Sec.~\ref{sec:mergers} (popIII, Q3,
Q3-nod) and seven values of $m_s$ in total, one for each decade in the
boson mass range
$m_s \in [10^{-19}, 10^{-13}]$~eV. 

\begin{figure}[t]
\begin{center}
\begin{tabular}{c}
\epsfig{file=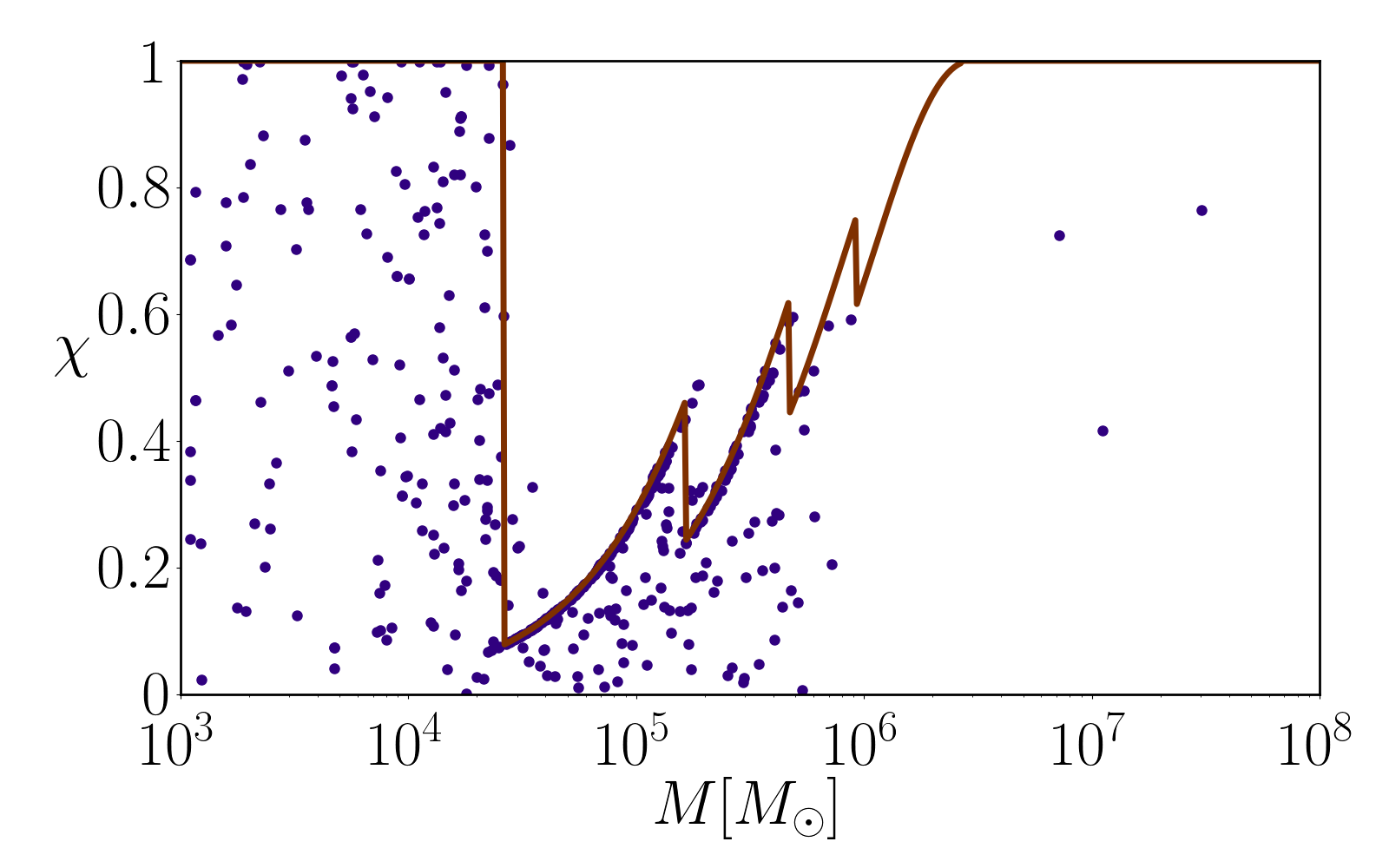,width=\columnwidth,angle=0,clip=true}
\end{tabular}
\caption{Example of a two-year simulation of massive BHs as observed
  by LISA assuming the Q3-nod model in the presence of a boson of mass
  $m_s = 10^{-16}$~eV. Each blue circle corresponds to the mass and
  spin of one component of an observed BH binary. The brown line
  corresponds to the maximum allowed spin $\chi_{\rm max}(M,\,m_s)$
  for the given boson mass. This curve is shaped like a sawtooth
  because different $m$-harmonics are more important for different BH
  masses. In this particular instance, LISA measurements from the
  simulated data would lead to a measured boson mass
  $0.88 \times 10^{-16}$~eV $< m_s^{\rm m} < 1.35 \times 10^{-16}$~eV.}
\label{fig:MBHB-obs-example}
\end{center}
\end{figure}
 
To simulate the loss of mass and angular momentum for each BH in the
catalogs we compute the final angular momentum $J_f$ and mass $M_f$
according to Eqs.~(\ref{finalspin}) and (\ref{finalmass}), with
azimuthal number $1 \leq m \leq 4$ and frequency given by~\eqref{omega}
with $l=m$ and $n=0$. Approximating $\omega_R \approx \mu$ in
Eq.~\eqref{omega} (which is strictly valid if $M\mu \ll 1$, but which is a
good approximation even for $M\mu$ of order unity) we get
\begin{align}
 \chi_f &= \frac{4 M_i \mu \left( m - M_i \mu \chi_i\right)}{m^2}, \\
 M_f &= \frac{m - \sqrt{m^2 - 4 m M_i \mu \chi_f + 4 M_i^2 \mu^2 
\chi_i \chi_f}}{2\mu\chi_f}.
\end{align}

We migrate BHs in the Regge plane if the age of the Universe $t(z)$ at the
merger redshift is larger than the instability time scale
($t(z) > \tau_{\rm inst} = 1/\omega_I$) and if the spin is higher than
a threshold $\chi_{\rm max} (M, m_s)$ set by
Eq.~\eqref{finalspin}. This migration causes BHs in the catalog to
accumulate along the critical line $\chi_{\rm max}(M,\,m_s)$ in the
Regge plane. An example of this accumulation can be seen in
Fig.~\ref{fig:MBHB-obs-example}.

To compare two models $\mathcal{M}_1$ and $\mathcal{M}_2$ given a set
of observations (i.e., a data set $D$), we can use Bayes' theorem. The
probability of model $\mathcal{M}_i$ given the observations is
\begin{align}
 P(\mathcal{M}_i | D) &= \frac{P(D | \mathcal{M}_i) P(\mathcal{M}_i)}{P(D)},
\end{align}
where $P(\mathcal{M}_i)$ is the prior on model $\mathcal{M}_i$,
$P(D | \mathcal{M}_i)$ is the likelihood of the data given the model,
and $P(D)$ is an overall probability of observing the data $D$. Given
a likelihood function for each model, we can then compute the odds
ratio between the two models:
\begin{align}
 \mathcal{O}( \mathcal{M}_1 / \mathcal{M}_2 ) &= \frac{P(\mathcal{M}_1 | 
D)}{P(\mathcal{M}_2 | D)} = \frac{P(D | \mathcal{M}_1)}{P(D | 
\mathcal{M}_2)} \frac{P(\mathcal{M}_1)}{P(\mathcal{M}_2)}.
\end{align}
A value of the odds ratio larger than one favors model
$\mathcal{M}_1$, while a value of the odds ratio lower than one favors
model $\mathcal{M}_2$. When $P(\mathcal{M}_1) = P(\mathcal{M}_2)$ the
last factor on the right-hand side simplifies, and the odds ratio is
just the ratio of the likelihood of the data in both models (also
known as the ``Bayes factor'').

We construct a likelihood function for BHs in the Regge plane for two
models: one with no ultralight boson, and one with an ultralight boson
of mass $m_s$. To avoid a possible bias toward high spins in the
astrophysical models (see e.g.~\cite{Sesana:2014bea}) we choose the
simplest likelihood function in the absence of bosons:
$\mathcal{L}_0(M, \chi) = 1$. In the presence of bosons, we set the
likelihood $\mathcal{L}_{m_s} (M, \chi)$ to unity if
$\chi \le \chi_{\rm max}(M,m_s)$, and we set it to zero otherwise. We
add to this likelihood a Gaussian centered on the threshold
$\chi_{\rm max}(M,m_s)$ with width $\sigma_\chi = 0.05$, with a
prefactor $1 - \chi_{\rm max}$ in front of it. This factor represents
the fraction of BHs with spins higher than the threshold that have
migrated out of the exclusion region to accumulate on the threshold
line, under the simplifying assumption that they migrate in the $\chi$
direction only (i.e., we neglect the relatively small variations in
the BH mass). In summary, the likelihood $\mathcal{L}_{m_s} (M, \chi)$
in the presence of a boson of mass $m_s$ is defined by
\begin{align}
 \mathcal{L}_{m_s} (M, \chi) &= \left\{
 \begin{array}{r r}
  1, & \chi_{\rm max} (M, m_s) = 1 \\
  1 + G(\chi, 0.05), & \chi < \chi_{\rm max} (M, m_s) < 1 \\
  G(\chi, 0.05), & \chi_{\rm max} (M, m_s) < \chi < 1 
 \end{array}
 \right. , \label{eq:likelihood-boson}\\
 G(\chi, \sigma) &= \frac{1 - \chi_{\rm max}}{\sqrt{2 \pi} \sigma} \exp\left[ 
-\frac{( \chi - \chi_{\rm max})^2}{2 \sigma^2} \right] .
\end{align}
The prefactor in front of the Gaussian ensures that the two
likelihoods $\mathcal{L}_0 (M, \chi)$ and $\mathcal{L}_{m_s}(M, \chi)$
have the same ``weight'', in the sense that the integral
$\int \mathcal{L}_{\mathcal{M}} dM d\chi$ is independent of the model
(so the presence or absence of an ultralight boson have, a priori, the
same probability). This choice for the likelihood functions assures that the computation of the odds ratio is agnostic about the underlying astrophysical model.

As stated earlier, the spin threshold $\chi_{\rm max} (M, m_s)$ is
given by Eq.~\eqref{finalspin}. In practice this criterion is slightly
complicated by the fact that the range of affected BH masses depends
on the time available for each system to radiate, which in turn
depends on the redshift. For simplicity we compute the spin limit
using a constant instability time scale of $500$~Myrs (approximately
the age of the Universe at redshift $z=10$), setting $\omega_R = \mu$
in Eq.~\eqref{finalspin}. The choice of this time scale is
conservative in the sense that the exclusion region is smaller than it would 
have been
if we had chosen longer time scales. Indeed, our choice reduces the likelihood 
discrepancy for low redshift BHs that will have migrated to the threshold line, 
but would not have had the time to do so had they merged at higher redshifts. 
For illustration, Fig.~\ref{fig:MBHB-obs-example} shows the
distribution of BH masses and spins for one realization of a two-year
catalog with $m_s = 10^{-16}$~eV, along with the corresponding spin
threshold $\chi_{\rm max} (M, m_s)$.

We simulate LISA observations of these catalogs using a Fisher-matrix
analysis similar to the study presented in~\cite{Klein:2015hvg}, using
the updated LISA noise PSD of~\cite{Audley:2017drz}. In addition to
instrumental noise, we also include the boson mass-dependent confusion
noise coming from superradiant BH instabilities shown in Fig.~1 of~\cite{Brito:2017wnc}. For each detectable binary (where detectability is
defined as $\rho>10$)\footnote{Note that this threshold is slightly different from that
used elsewhere in the paper ($\rho=8$, though that was for boson-condensate sources). Still, 
the results hardly depend on this choice, since barely detectable events ($\rho\sim 8-10)$ have anyhow very poor spin determinations.}
we approximate the recovered distribution for
each binary BH component by a bivariate Gaussian centered on the true
values $(\bar{M}_i, \bar{\chi}_i)$, with spread given by the
two-dimensional inverse of the covariance matrix
$\Gamma = \Sigma^{-1}$:
\begin{multline}
 p_{\rm obs}(M_i, \chi_i) = \frac{\sqrt{|\Gamma|}}{2 \pi} \exp\bigg\{ 
-\frac{1}{2} \big[
\Gamma_{M_i M_i} (M_i - \bar{M}_i)^2 \\
+ \Gamma_{\chi_i \chi_i} (\chi_i - 
\bar{\chi}_i)^2 + 2 \Gamma_{M_i \chi_i} (M_i - \bar{M}_i)(\chi_i - 
\bar{\chi}_i) 
\big]\bigg\}.\label{pobs}
\end{multline}

One problem is that GW observations can measure the reshifted mass
$M_z = (1 + z) M$, rather than the BH mass in the source frame
$M$. Lensing effects will induce an extra uncertainty on the distance
to the source of typical size $\sigma_{D_L}^{\rm lens}(z)$, and
through the redshift-distance relation $D_L(z)$ an extra uncertainty
on the redshift of size $\sigma_z^{\rm lens}(z)$. We include the
effects of lensing by adjusting the observed distribution
$p_{\rm obs}(M_i, \chi_i)$ along the mass direction.  We estimate the
typical extra error on the mass due to lensing as
\begin{align}
 \frac{\sigma_M^{\rm lens}(z)}{M} &= \frac{\sigma_z^{\rm lens}(z)}{1 + z} = 
\frac{dz}{dD_L}(z) \frac{\sigma_{D_L}^{\rm lens}(z)}{1 + z}.
\end{align}
where the luminosity distance error as a function of redshift can be
estimated by the approximate
relation~\cite{Hirata:2010ba,Tamanini:2016}
\begin{align}
 \sigma_{D_L}^{\rm lens}(z) &= D_L(z) \times 0.066 \left\{4 \left[1 - 
(1+z)^{-1/4} \right] \right\}^{9/5}.
\end{align}

At this stage we can compute the likelihood of an observed BH for each
model $\mathcal{M}$ by integrating the product
\begin{align}
 \mathcal{L}(i | \mathcal{M}) = \int p_{\rm obs}(M_i, \chi_i) 
\mathcal{L}_\mathcal{M} (M_i, \chi_i) dM_i d\chi_i, \label{eq:likelihood}
\end{align}
where the index $i$ labels the observed BH. In the absence of
ultralight bosons we get $\mathcal{L}(i|\mathcal{M}_0) = 1$, and in
the presence of bosons we use Monte Carlo methods to compute
$\mathcal{L}(i|m_s)$. In practice we generate a set of random points
in the Regge plane $(M_k, \chi_k)$ distributed according to
$p_{\rm obs}(M_i, \chi_i)$, with an extra (spin-independent) jump in
the mass direction due to lensing, which we assume to be Gaussian
distributed with zero mean and standard deviation
$\sigma_M^{\rm lens}(z)$. The integral is then approximated by
\begin{align}
 \mathcal{L}(i | m_s) &\approx \frac{1}{N} \prod_{k=1}^{N} 
\mathcal{L}_{m_s}(M_k, \chi_k).
\end{align}
The integration with respect to mass and spin in Eq.~\eqref{eq:likelihood} 
tends to suppress the effect on the odds ratio of potential
observations in the exclusion region that would favor high spins with low 
confidence. As one can see from Eq.~\eqref{pobs}, if the measurement error on the spin is significant, 
Eq.~\eqref{eq:likelihood} will show a significant overlap between the two 
factors inside the integral, even if the observed spin is higher than the 
threshold.

Using this method we can simulate a set of LISA observations $D$ and
compute its likelihood for model $\mathcal{M}$ as
\begin{align}
 \mathcal{L} ( D | \mathcal{M} ) = \prod_i \mathcal{L}(i|\mathcal{M}),
\end{align}
where the product is taken over all components of a binary observed
with SNR $\rho>10$.  Then, assuming no prior preference, we compute the
odds ratio between a model with boson mass $m_s$ and a model without
bosons:
\begin{align}
 \mathcal{O} ( m_s / \mathcal{M}_0 ) &= \mathcal{L} ( D | m_s).
\end{align}

We simulated
observations in the absence of ultralight bosons and in the presence
of an ultralight boson with seven possible values of $m_s$ (one for each
decade in the boson mass range $m_s \in [10^{-19}, 10^{-13}]$~eV).
For each boson mass and for the model without bosons, we simulated a 
set of 21 realizations of the LISA mission considering the three
astrophysical models (popIII, Q3, and Q3-nod) and four choices for the
observation time (6 months, 1 year, 2 years and 4 years), corresponding to 
a total of 252 simulations per model.

\begin{figure}[t]
\begin{center}
\begin{tabular}{c}
\epsfig{file=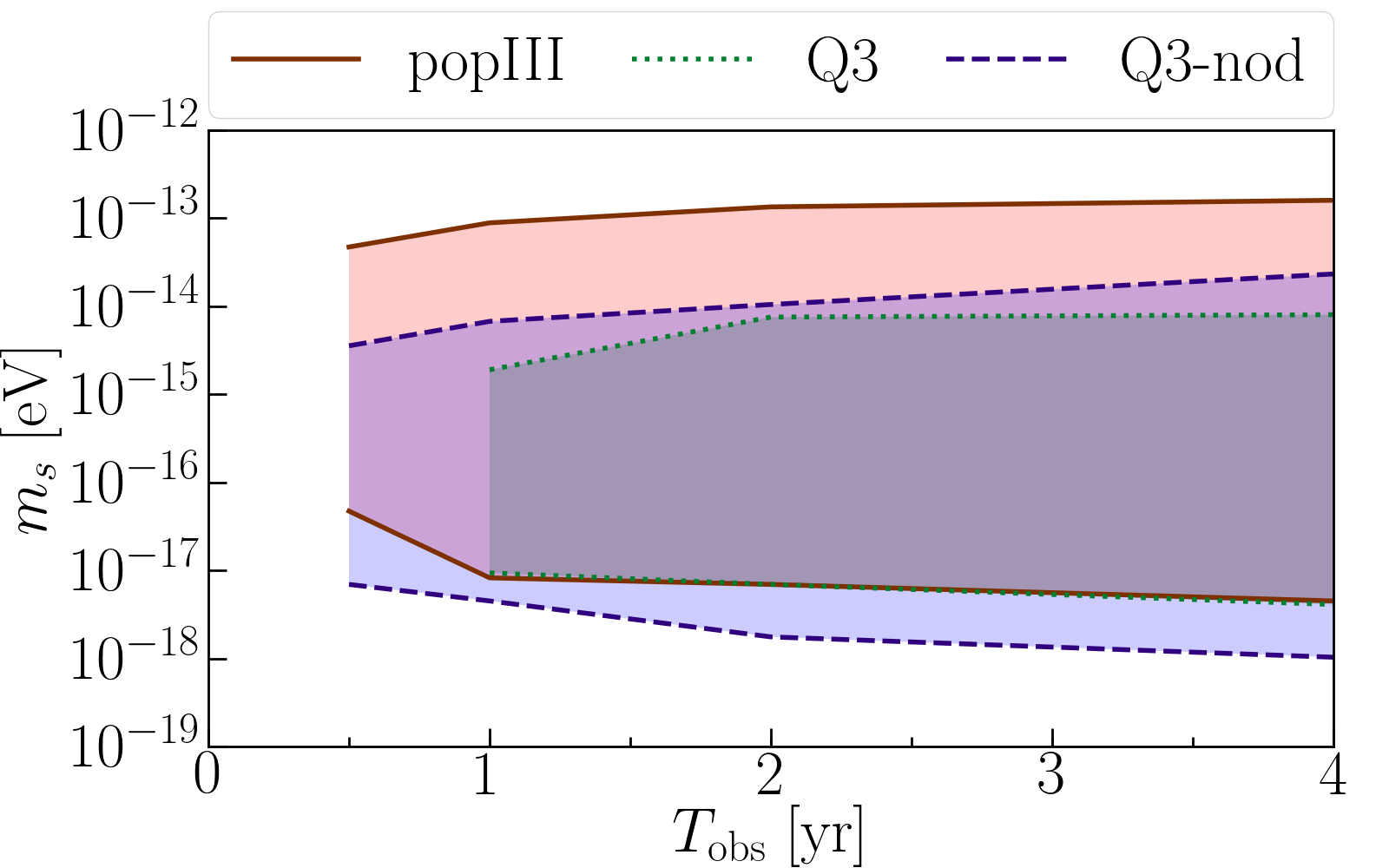,width=8cm,angle=0,clip=true}
\end{tabular}
\caption{Median minimum and maximum boson mass excluded by LISA for
  different observation times $T_{\rm obs}$ and BH evolution models
  (red, solid line: popIII; green, dotted line: Q3; blue, dashed line:
  Q3-nod). Due to the reduced merger rate in the Q3 model, limits on the boson 
mass could be put in more than half of the simulations only after one year of 
observations.}
\label{fig:MBHB-exclusion}
\end{center}
\end{figure}

In the absence of an ultralight boson, we identify as excluded the range of 
masses where the odds ratio $\mathcal{O}$ satifies
$\log[\mathcal{O}(m_s /\mathcal{M}_0)] < -4.5$. This criterion
corresponds to rejecting the presence of the given boson mass at
3-$\sigma$ confidence level. This requirement to exclude a boson of a
given mass corresponds to a false alarm rate of $\sim 10\%$ for a four-year 
mission
in the popIII model, and less than $5\%$ in the other models: the 
maximum odds ratio incorrectly favoring the
presence of an ultralight boson in the 84 realizations where we assumed 
its
absence was $\log( \mathcal{O}_{\rm max}) = 5.2$ for the popIII model, $1.1$ 
for the Q3 model, and $2.8$ for the Q3-nod model. In the popIII case, a maximum 
odds ratio of $4.5$ was exceeded twice.  The median range of
boson masses excluded in our simulations is summarized in
Table~\ref{tab:MBHB-exclusion} and Fig.~\ref{fig:MBHB-exclusion}. As
expected, in our light-seed (popIII) model the excluded boson masses
are higher than in the heavy-seed models Q3 and Q3-nod, because the
observed BH masses are generally lower in light-seed scenarios.  The
Q3-nod model allows us to set more stringent bounds than the Q3 model,
because the merger rate is higher when there are no delays between
galaxy mergers and BH mergers. Furthermore, the Q3 model failed to allow for a 
boson mass exclusion after six months of observations in 12 of the 21
simulations due to its low merger rate. For any astrophysical models among the
three we considered, four years of LISA observations would allow us to
exclude boson masses in the range $4.1 \times 10^{-18}$~eV to
$8 \times 10^{-15}$~eV.

\begin{table}[h]
\begin{center}
\begin{tabular}{cc|cc}
Model & $T_{\rm obs}$ [yr] & Min [eV] & Max [eV]\\
\hline 
\hline
popIII   &  0.5  &  $4.7 \times 10^{-17}$  &  $4.7 \times 10^{-14}$\\
   &  1  &  $8.2 \times 10^{-18}$  &  $8.9 \times 10^{-14}$\\
   &  2  &  $6.9 \times 10^{-18}$  &  $1.3 \times 10^{-13}$\\
   &  4  &  $4.5 \times 10^{-18}$  &  $1.6 \times 10^{-13}$\\
\hline
Q3   &  0.5  &  --  &  --\\
   &  1  &  $9.4 \times 10^{-18}$  &  $1.9 \times 10^{-15}$\\
   &  2  &  $6.9 \times 10^{-18}$  &  $7.5 \times 10^{-15}$\\
   &  4  &  $4.1 \times 10^{-18}$  &  $8 \times 10^{-15}$\\
\hline
Q3-nod   &  0.5  &  $6.9 \times 10^{-18}$  &  $3.6 \times 10^{-15}$\\
   &  1  &  $4.5 \times 10^{-18}$  &  $6.7 \times 10^{-15}$\\
   &  2  &  $1.8 \times 10^{-18}$  &  $1 \times 10^{-14}$\\
   &  4  &  $1 \times 10^{-18}$  &  $2.3 \times 10^{-14}$
\end{tabular}
\end{center}
\caption{Median minimum and maximum boson mass excluded by LISA 
  for different BH evolution models (popIII, Q3, Q3-nod) and
  observation times $T_{\rm obs}$.}
\label{tab:MBHB-exclusion}
\end{table}

\begin{table*}[t]
\begin{center}
\begin{tabular}{cc|cc|cc|cc|cc}
& $T_{\rm obs}$ [yr] &
\multicolumn{2}{c|}{0.5} & \multicolumn{2}{c|}{1} & \multicolumn{2}{c|}{2} & 
\multicolumn{2}{c}{4}  \\
Model & $m_s$~[eV] &
$\kappa$ & $L$ &
$\kappa$ & $L$ &
$\kappa$ & $L$ &
$\kappa$ & $L$  \\
\hline
\hline
popIII & $10^{-16}$ &  --  & 2.9
  & -- & 4.1
   &  $1.06 \pm 0.25$  & 13 
&  $1.07 \pm 0.12$  & 28 \\
& $10^{-15}$ &  $1 \pm 0.4$  & 7.9
   & $1.05 \pm 0.21$ & 14
   &  $1.06 \pm 0.11$  & 39
   &  $1.08 \pm 0.06$  & 90 \\
& $10^{-14}$ &  $1 \pm 0.6$  & 5.4 
 & $1.02 \pm 0.15$ & 12
   &   $1.05 \pm 0.1$  & 31
   &  $1.06 \pm 0.06$  & 81 \\
& $10^{-13}$ & -- & 0.64 & -- & 1.7 & $1 \pm 0.15$ & 
8.6 
& $1.02 \pm 0.1$ & 26 \\
\hline
Q3 & $10^{-16}$  &  --  & 0.91 
   &  -- & 3.2 
   &   --  & 4.5 
   &    $1 \pm 0.4$  & 9.7 \\
   & $10^{-15}$  &  --  & 0 
   &  -- & 1.9
   &   --  & 3.6 
   &    $1 \pm 0.4$  & 6.8 \\
\hline
Q3-nod & $10^{-17}$ &  --  & 2.9 
   &   $1 \pm 0.23$ & 6.5 
   &    $1.03 \pm 0.19$  & 13
   &    $1.02 \pm 0.13$  & 25 \\
& $10^{-16}$ &  $1 \pm 0.4$  & 17 
   &   $0.99 \pm 0.15$ & 47 
   &    $1 \pm 0.08$  & 98 
   &    $0.97 \pm 0.06$  & 200 \\
& $10^{-15}$ &  $1 \pm 0.5$  & 11 
   &   $0.94 \pm 0.18$ & 28 
   &    $0.95 \pm 0.1$  & 65 
   &   $0.98 \pm 0.07$  & 140  \\
& $10^{-14}$ &  -- & 1.6 
   &   -- & 4.2 
   &    $0.98 \pm 0.21$  & 14 
   &    $0.98 \pm 0.13$  & 27 
\end{tabular}
\end{center}
\caption{Median measured boson mass $m_s^{\rm m} = \kappa m_s$ and median 
maximum 
  log likelihood $L = \log \mathrm{O}_{\rm max}$ for different BH evolution models, 
  observation times $T_{\rm obs}$, and ``true'' boson masses $m_s$.}
\label{tab:MBHB-measurement}
\end{table*}

It is also interesting to address the following question: in the presence of
an ultralight boson, could it be detected? And if so, what is the
accuracy with which we could determine its mass? To answer the
first question we identify the mass range where
$\log[\mathcal{O}(m_s) / \mathcal{O}_{\rm max}] \geq -4.5$, again
corresponding to a 3-$\sigma$ confidence level, and then use the
simulated events to determine the accuracy with which $m_s$ can be
determined. Our results are summarized in
Table~\ref{tab:MBHB-measurement}. We do not show results for masses where four 
years of observations were not enough to claim a boson detection. In marginal 
detections
($\log(\mathcal{O}_{\rm max}) \lesssim 10$), only the order of
magnitude of the boson mass could be inferred.

For the light-seed popIII model, boson masses in the range
$[10^{-16}, 10^{-13}]$~eV could be confidently detected after four
years of observations with measurement errors in $m_s$ of  $5$-$10$~\%.
Model Q3-nod allows for the confident detection of a
boson in the mass range $[10^{-17}, 10^{-14}]$~eV with mass
measurement errors of $5$-$15$~\%, while the less optimistic model
Q3 only allows detections for bosons with mass in the range $[10^{-16}, 
10^{-15}]$~eV, with
 mass measurement errors of $\sim40 \%$.  

We remark that the biases in the recovered boson masses are sometimes
comparable to the corresponding measurement accuracies: in low-mass
(high-mass) seed models we tend to overestimate (underestimate) the
boson mass.
It is likely that this bias could be reduced with better modeling of
the relevant physics -- e.g. by evolving Eqs.~\eqref{evolution}
numerically for each BH from formation until merger -- or with a more
careful choice of the likelihood function, e.g. by taking the observed 
redshift of the system into account in the definition of the threshold line in 
Eq.~\eqref{eq:likelihood-boson}, i.e. in the likelihood in the presence of 
bosons.  A more detailed analysis of
systematic and statistical errors in recovering the boson masses is an
interesting topic for future work.

\section{Conclusions and outlook}

In this work and in the companion paper~\cite{Brito:2017wnc} we assess the
detectability of light-boson condensates around BHs with GW
interferometers combining the best available estimates for GW emission
from these systems, state-of-the-art astrophysical BH population
models, and relatively realistic GW data analysis techniques.

For both Advanced LIGO and LISA, we find that the most stringent
constraints on the boson mass $m_s$ should come from the stochastic
background produced by the superposition of unresolved GW signals from
BH-boson condensate systems.  We show that this background should be
detectable by Advanced LIGO for
$m_s\in [2\times 10^{-13}, 10^{-12}]\,{\rm eV}$, and by LISA for
$m_s\in [5\times 10^{-19}, 5\times 10^{-16}]\,{\rm eV}$.  We also find
that existing constraints on the stochastic background from Advanced
LIGO's O1 run may {\em already} rule out a range of boson masses in
the Advanced LIGO window. Although the precise constrained regions depend on the astrophysical model, the order of magnitude of the stochastic background is robust with respect to astrophysical
 uncertainties, as shown in Fig. 2 of~\cite{Brito:2017wnc}.

Our results indicate that $\sim 15-200$ resolvable sources should be
detectable by Advanced LIGO for scalar field masses
$m_s\sim3\times 10^{-13}$~eV, while LISA should be able to resolve
$\sim5-40$ sources for $m_s\sim 10^{-17}$~eV.  Moreover, LISA
measurements of BH spins may either determine
$m_s\in [10^{-17}, 10^{-13}]$~eV to within $10\%$ accuracy, or rule
out boson masses in the range
$m_s\in [10^{-18}, 1.6\times 10^{-13}]$~eV. 

We anticipate that pulsar-timing
arrays~\cite{2013CQGra..30v4009K,2013CQGra..30v4007H,2013CQGra..30v4008M,2010CQGra..27h4013H,2013CQGra..30v4010M},
though sensitive to the stochastic GW background in the nHz band, may
not set stringent constraints on the masses of ultralight bosons. The
reason lies in the very large instability and gravitational radiation
time scales for bosons masses in the nHz band and in the paucity of
massive BHs with masses
$M\gtrsim 10^{10} M_\odot$~\cite{Natarajan:2008ks,Pacucci:2017zgh},
which would be required to produce a significant background from
BH-boson condensates.  Conversely, an interferometer like
DECIGO~\cite{Kawamura:2006up} would allow one to put constraints on
boson masses $m_s\sim 10^{-14}$~eV.

Some of our conclusions differ from previous work on this topic by
Arvanitaki {\it et al.}~\cite{Arvanitaki:2014wva,Arvanitaki:2016qwi},
which neglected the stochastic background from boson condensates in
the LISA and in the LIGO band, focusing on resolved events. This had
the two-fold effect of (i) missing the strong constraints (summarized
above) from existing and projected stochastic background limits, and
(ii) missing the ``self-confusion'' problem, i.e. the fact that the
stochastic background itself is a confusion noise (similar to the
familiar white dwarf confusion noise in the LISA band), impairing the
detectability of individual sources.

Another important difference with respect to Arvanitaki et
al.~\cite{Arvanitaki:2014wva,Arvanitaki:2016qwi} lies in our
astrophysical
models. Refs.~\cite{Arvanitaki:2014wva,Arvanitaki:2016qwi} focused on
Galactic BHs as resolvable LIGO sources. This is probably the main
reason why they overlooked the presence of a significant stochastic
background, which is mostly produced by extragalactic BHs. Likewise,
the lower LISA event rates found by \cite{Arvanitaki:2014wva} (in
spite of their neglecting the aforementioned confusion noise from the
background) seem to be due to their simplified (and overly
pessimistic) models for the massive BH population. For example, Ref.~\cite{Arvanitaki:2014wva} considered the chaotic accretion model of~\cite{King:2008au}, where BHs with large spins are unlikely. Such models are either disfavored or ruled out (depending on the assumed spin distribution) by iron K$\alpha$ line data~\cite{Sesana:2014bea}.

Finally, our analysis of the statistical error affecting GW
measurements of BH spins in the LISA band and our use of Bayesian
model selection techniques (while far from realistic) are a step
forward with respect to the estimates of~\cite{Arvanitaki:2016qwi},
and they lead to one of the most remarkable conclusions of our
work. As shown schematically in Fig.~\ref{fig:Regge}, LISA could
either rule out light bosons in the mass range
$[4 \times 10^{-18}, 10^{-14}]$~eV, or measure $m_s$ with ten percent
accuracy if particles in the mass range $[10^{-17}, 10^{-13}]$~eV
exist in Nature.

\noindent{\bf{\em Acknowledgments.}}
%
%
We thank Asimina Arvanitaki, Masha Baryakhtar and Robert Lasenby for useful comments.
S.~Ghosh and E.~Berti are supported by NSF Grants No.~PHY-1607130,
AST-1716715 and by FCT contract IF/00797/2014/CP1214/CT0012 under the
IF2014 Programme.
V.~Cardoso acknowledges financial support provided under the European Union's H2020 ERC Consolidator Grant ``Matter and strong-field gravity: New frontiers in Einstein's theory'' grant agreement no. MaGRaTh--646597.
The work of I.~Dvorkin
has been done within the Labex ILP (reference ANR-10-
LABX-63) part of the Idex SUPER, and received financial
state aid managed by the Agence Nationale de la Recherche,
as part of the programme Investissements d'avenir under the
reference ANR-11-IDEX-0004-02.
Research at Perimeter Institute is supported by the Government of Canada through Industry Canada and by the Province of Ontario through the Ministry of Economic Development $\&$
Innovation.
This project has received funding from the European Union's Horizon 2020 research and innovation programme under the Marie Sklodowska-Curie grant agreement No 690904 and from FCT-Portugal through the project IF/00293/2013.
The authors would like to acknowledge networking support by the COST Action CA16104.
This work has made use of the Horizon Cluster, hosted by the Institut
d'Astrophysique de Paris. We thank Stephane Rouberol for running
smoothly this cluster for us.
The authors thankfully acknowledge the computer resources, technical expertise and assistance provided by S\'ergio Almeida at CENTRA/IST. Computations were performed at the cluster ``Baltasar-Sete-S\'ois'', and supported by the MaGRaTh--646597 ERC Consolidator Grant.
This work was supported by the French Centre National d'{\'E}tudes Spatiales 
(CNES).
%


\bibliographystyle{apsrev4}
\bibliography{Ref}

\end{document}